\documentclass[aps,prd,reprint,floatfix,nofootinbib,nolongbibliography]{revtex4-2}

\usepackage[T1]{fontenc}
\usepackage[utf8]{inputenc}
\usepackage{graphicx}
\usepackage[usenames,dvipsnames]{xcolor}
\usepackage{booktabs}
\usepackage{mathtools}
\usepackage{amssymb}
\usepackage{lmodern}
\usepackage{siunitx}
\usepackage{slashed}
\usepackage[final]{microtype}
\usepackage{hyperref}
 
\hypersetup{%
	pdftitle = {Quarks and light (pseudo-)scalar mesons at finite chemical potential},
	pdfauthor = {Pascal J.~Gunkel, Christian S.~Fischer, Philipp Isserstedt},
	bookmarksopen = true,
	colorlinks = true,
	linkcolor = blue!50!black,
	citecolor = blue!50!black,
	urlcolor = blue!50!black
}

\newcolumntype{d}[1]{D{.}{.}{#1}}

\newcommand*{\+}{\hspace*{.08335em}}
\newcommand*{\?}{\hspace*{-.08335em}}
\newcommand*{\beq}{\begin{equation}}
\newcommand*{\eeq}{\end{equation}}
\newcommand*{\Nf}{N_{\textup{f}}}
\newcommand*{\Nc}{N_{\textup{c}}}
\newcommand*{\OO}{\textup{O}}
\newcommand*{\ZZ}{\textup{Z}}
\newcommand*{\dd}{\textup{d}}
\newcommand*{\upB}{\textup{B}}
\newcommand*{\upu}{\textup{u}}
\newcommand*{\upd}{\textup{d}}
\newcommand*{\ups}{\textup{s}}
\newcommand*{\upq}{\textup{q}}
\newcommand*{\ZT}{Z_{\textup{T}}}
\newcommand*{\ZL}{Z_{\textup{L}}}
\newcommand*{\muq}{\mu_{\upq}}
\newcommand*{\muB}{\mu_{\upB}}

\DeclareMathOperator*{\SumInt}{%
\mathchoice%
  {\ooalign{$\displaystyle\sum$\cr\hidewidth$\displaystyle\int$\hidewidth\cr}}
  {\ooalign{\raisebox{.14\height}{\scalebox{.7}{$\textstyle\sum$}}\cr\hidewidth$\textstyle\int$\hidewidth\cr}}
  {\ooalign{\raisebox{.2\height}{\scalebox{.6}{$\scriptstyle\sum$}}\cr$\scriptstyle\int$\cr}}
  {\ooalign{\raisebox{.2\height}{\scalebox{.6}{$\scriptstyle\sum$}}\cr$\scriptstyle\int$\cr}}
}

\DeclareMathOperator{\Tr}{Tr}
\DeclarePairedDelimiterX{\expval}[1]{\langle}{\rangle}{#1}

\DeclareSIUnit{\eV}{\electronvolt}

\begin{document}

\title{Quarks and light (pseudo-)scalar mesons at finite chemical potential}

\author{Pascal J.~Gunkel}
\email{pascal.gunkel@physik.uni-giessen.de}
\affiliation{%
	Institut f\"ur Theoretische Physik, %
	Justus-Liebig-Universit\"at Gie\ss{}en, %
	35392 Gie\ss{}en, %
	Germany%
}

\author{Christian S.~Fischer}
\email{christian.fischer@theo.physik.uni-giessen.de}
\affiliation{%
	Institut f\"ur Theoretische Physik, %
	Justus-Liebig-Universit\"at Gie\ss{}en, %
	35392 Gie\ss{}en, %
	Germany%
}

\author{Philipp Isserstedt}
\email{philipp.isserstedt@physik.uni-giessen.de}
\affiliation{%
	Institut f\"ur Theoretische Physik, %
	Justus-Liebig-Universit\"at Gie\ss{}en, %
	35392 Gie\ss{}en, %
	Germany%
}

\date{\today}

\begin{abstract}
We investigate the properties of light scalar and pseudoscalar mesons at finite (light) quark chemical potential. To this end we solve a coupled set of (truncated) Dyson-Schwinger equations for the quark and gluon propagators in Landau-gauge QCD and extent earlier results for $\Nf=2+1$ dynamical quark flavors to finite chemical potential at zero temperature. We then determine the meson bound state masses, wave functions, and decay constants for chemical potentials below the first-order phase transition from their homogeneous Bethe-Salpeter equation. We study the changes in the quark dressing functions and Bethe-Salpeter wave functions with chemical potential. In particular, we trace charge-conjugation parity breaking. Furthermore, we confirm the validity of the Silver-Blaze property: all dependencies of colored quantities on chemical potential cancel out in observables and we observe constant masses and decay constants up to and into the coexistence region of the first-order chiral phase transition.
\end{abstract}

\maketitle

\section{\label{sec:intro}Introduction}

Exploring the QCD phase diagram and the associated properties of hadrons at finite chemical potential 
is a difficult endeavour. At zero chemical potential, results from lattice QCD
\cite{Borsanyi:2010bp,Bazavov:2011nk,Bhattacharya:2014ara,Bazavov:2014pvz}
find an analytic crossover from a low-temperature phase characterized by confinement and 
chiral symmetry breaking to a high-temperature deconfined and (partially) chirally 
restored phase where the quark-gluon plasma  is realized. At large chemical potential, however,
lattice QCD is plagued by the fermion sign problem and one has to resort to functional methods and 
model studies, see, e.g., Refs.~\cite{Drews:2016wpi,Fukushima:2017csk,Fischer:2018sdj} for recent 
review articles and comprehensive guides to the literature. 

At zero temperature and small chemical potential, the so-called Silver-Blaze property of QCD is at work: 
before the baryon chemical potential $\muB$ is large enough to create physical excitations 
(i.e., for $\muB$ smaller than the mass of the nucleon minus its binding energy in nuclear matter), 
the system has to stay in its vacuum ground state. This can be shown analytically for the case of 
finite isospin chemical potential, but is also extremely plausible for the case of finite baryon 
chemical potential \cite{Cohen:1991nk,Cohen:2004qp} and has been demonstrated for heavy quark masses 
in the lattice formulation of  Ref.~\cite{Fromm:2012eb}. As a consequence, all observables such as meson masses 
and decay constants have to remain constant in the Silver-Blaze region.

The physics of the QCD phase diagram is connected with the properties of mesons at finite temperature and 
chemical potential in various ways. When it comes to dynamical chiral symmetry breaking and its restoration, 
pseudoscalar and scalar mesons play an important role. The pseudoscalar mesons are the (pseudo-)Goldstone 
bosons of chiral symmetry breaking and the scalar mesons are their chiral partners. They constitute 
the effective long-range degrees of freedom that control the universal behavior of QCD at points of 
second-order chiral phase transitions. 
In the chiral limit of the two-flavor theory (and depending on the fate of the $\textup{U}_{\textup{A}}(1)$-symmetry) we may 
encounter a second-order transition in the $\OO(4)$ universality class with the isovector pions and the 
isoscalar scalar meson as driving degrees of freedom. At a putative critical end point for physical quark 
masses, this may turn into the $\ZZ(2)$ universality class driven by the isoscalar scalar meson. 

Mesons are therefore interesting objects to study at finite temperature and chemical potential, see, e.g.,
Refs.~\cite{Rapp:1999ej,Leupold:2009kz} for reviews. On the level of effective theories, the quark-meson model exploits the above mentioned chiral physics and is a convenient tool for many applications associated
with the QCD phase diagram and its relations to heavy-ion collisions \cite{Fukushima:2017csk}. This is
particularly true for versions adding the Polyakov-loop potential and using the functional renormalization
group, see, e.g., Refs.~\cite{Skokov:2010wb,Herbst:2010rf,Herbst:2013ail,Tripolt:2017zgc,Resch:2017vjs}
and references therein. 

A very interesting aspect of mesons is their nature as composite
particles bound by the forces of QCD. The dichotomous nature of the pion as a Goldstone boson and a bound 
state of a quark and an antiquark has been extensively studied in the vacuum, see, e.g., Ref.~\cite{Horn:2016rip} 
for a comprehensive review. In this work we extent the study of pions and sigma mesons to (zero temperature 
and) finite chemical potential in and beyond the Silver-Blaze region up to the first-order chiral phase 
transition within the framework of Dyson-Schwinger equations (DSEs). We complement and develop previous work 
on the topic \cite{Maris:1997eg,Bender:1997jf,Zong:2005mm,Jiang:2008zzd,Jiang:2008rb} using a well-studied
truncation scheme that includes the back-reaction of the quarks onto the gluons. 
Our aim is to carefully analyze the interplay of changes in the quark propagation as well as in the
Bethe-Salpeter wave function of the mesons that need to conspire such that the Silver-Blaze property 
remains satisfied. This may also provide useful information for intended later applications at high
temperature and chemical potential along the phase boundary of QCD up to and in the vicinity of
the critical end point. 

The paper is organized as follows: In Sec.~\ref{sec:dse_bse} we explain our setup for the truncation
of the Dyson-Schwinger equations for quarks and gluons and the associated interaction kernel in the 
meson Bethe-Salpeter equation. We go beyond previous works by (i) explicitly back-coupling the 
quarks onto the gluon, (ii) solving the quark DSE in the complex momentum plane providing the necessary
input to be able to, (iii), solve the Bethe-Salpeter equation explicitly thereby going beyond the leading
Dirac tensor approximation.  In Sec.~\ref{sec:results}, we discuss our results for quarks and mesons
at finite chemical potential. We conclude with a short summary and outlook in Sec.~\ref{sec:summary}.

\section{\label{sec:dse_bse}Dyson-Schwinger/Bethe-Salpeter formalism}

\subsection{\label{sec:quark_gluon_propagator}Quark and gluon propagators}

The dressed quark propagator $S_{f}$ for a quark flavor $f$ and the Landau-gauge gluon propagator $D_{\mu\nu}$ for finite quark chemical potential 
$\mu_{\textup{q}}^{f}$ but vanishing temperature $T$ are given 
by\footnote{We work in four-dimensional Euclidean space-time with Hermitian gamma matrices obeying
	$\{ \gamma_\mu, \gamma_\nu \} = 2 \+ \delta_{\mu\nu}$; $\mu,\nu \in \{ 1, 2, 3, 4 \}$.}
\begin{align}
S^{-1}_f(p)=&\;i\vec{\slashed{p}}\+A_f(p)+i\tilde{p}_{4}\gamma_{4}\+C_f(p)+B_f(p)\,, \\
D_{\mu\nu}(k)=&\;\mathcal{P}_{\mu\nu}^{\textup{T}}(k) \+ \frac{\ZT(k)}{k^{2}}+\mathcal{P}_{\mu\nu}^{\textup{L}}(k)\+\frac{\ZL(k)}{k^{2}}
\label{eq:dressed_quark_gluon_propagator_ansatz_landau_gauge}
\end{align}
with momenta $p=(\vec{p},\tilde{p}_{4})$ and $k=(\vec{k},k_{4})$ and the abbreviation
$\tilde{p}_{4}=p_{4}+i\mu_{\textup{q}}^{f}$. The dressing functions $A_f$, $B_f$, $C_f$ have a non-trivial momentum dependence and encode the non-perturbative information. 
The projectors $\mathcal{P}_{\mu\nu}^{\textup{T},\textup{L}}(k)$ are longitudinal (L) and transversal (T) 
to the assigned direction $v=(\vec{0},1)$ of the medium and defined by
\begin{align}
\mathcal{P}_{\mu\nu}^{\textup{T}}(k)=&\left(\delta_{\mu\nu}-\frac{k_{\mu}k_{\nu}}{\vec{k}^{2}}\right)(1-\delta_{\mu4})\,(1-\delta_{\nu4})\,, \\
\mathcal{P}_{\mu\nu}^{\textup{L}}(k)=&\;T_{\mu\nu}(k)-\mathcal{P}_{\mu\nu}^{\textup{T}}(k)\,, \\
T_{\mu\nu}(k)=&\;\delta_{\mu\nu}-\frac{k_{\mu}k_{\nu}}{k^{2}}\,.
\label{eq:gluon_projectors}
\end{align}
For vanishing chemical potential, the two quark vector dressing functions $A_{f}$ and $C_{f}$ as well 
as the two gluon dressing functions $\ZT$ and $\ZL$ are degenerate, whereas they become 
distinguishable in principle once chemical potential is switched on. 

\begin{figure}[t]
 \centering
 \includegraphics[scale=1,trim=16pt 0 0 0]{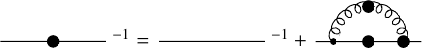}\\[1.5em]
 \includegraphics[scale=1]{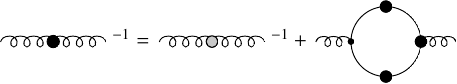}
 \caption{\label{fig:Quark_Gluon_DSE}%
 DSEs for the inverse quark (upper) and gluon (lower) propagator. Quark propagators are denoted by solid lines,
  while gluon propagators are represented as curly lines. Dressed quantities are denoted by big dots.}
\end{figure}

The DSEs for the quark and gluon propagator are shown diagrammatically in Fig.~\ref{fig:Quark_Gluon_DSE}.
In the gluon DSE the first term on the right hand side, marked with a grey dot, summarizes all 
diagrams already present in Yang-Mills theory. The remaining quark loop diagram couples the matter 
sector of QCD back onto the gluon. The diagram includes an implicit sum over all considered quark flavors. 
These equations are exact but need to be truncated in practice in order to be solved. 

In this work we are building upon the truncation of the gluon DSE and the quark-gluon vertex 
detailed in Ref.~\cite{Eichmann:2015kfa}. It evolved from the quenched case \cite{Fischer:2009wc,Fischer:2010fx}, 
to $\Nf=2$ \cite{Fischer:2011pk,Fischer:2011mz,Fischer:2012vc} 
and finally to $\Nf=2+1$ and $\Nf=2+1+1$ quark flavors \cite{Fischer:2012vc,Fischer:2014ata}. 
Applications also include color superconductivity \cite{Muller:2013pya,Muller:2016fdr}.
An overview can be found in the recent review Ref.~\cite{Fischer:2018sdj}.

Since details of the truncation are given in Ref.~\cite{Eichmann:2015kfa}, here we only summarize 
the most important aspects. We use a temperature-dependent fit $D_{\mu\nu}^{\textup{que.\,fit}}(k)$ 
to quenched lattice data \cite{Fischer:2010fx,Maas:2011ez} for the Yang-Mills diagrams in the gluon 
DSE and an ansatz for the quark-gluon vertex that is discussed below. The quark-loop diagram then 
effectively unquenches the gluon. A measure of the quality of this truncation can be obtained by
comparing the DSE results for the unquenched gluon \cite{Fischer:2013eca} with corresponding lattice 
results \cite{Aouane:2012bk}. Indeed, there is very good agreement. 

The system of coupled integral equations for the inverse quark and gluon propagators then reads explicitly
\begin{align}
S_{f}^{-1}(p)=&\;S_{f,0}^{-1}(p)+C_{\textup{F}} \+ g^{2} \+ Z_{\textup{1F}}^{f} \nonumber \\[0.25em]
& \phantom{S_{f,0}^{-1}(p)+\;} \times\int_{q} \, \gamma_{\mu} \+ S_{f}(q) \+ \Gamma_{\nu}^{f}(p,q,k) \+ D_{\mu\nu}(k)\,, \\[0.5em]
D_{\mu\nu}^{-1}(k)=&\;\left[D_{\mu\nu}^{\textup{que.\,fit}}(k)\right]^{-1} \label{eq:quark_gluon_DSE} \nonumber \\[0.25em]
&\;-\frac{g^{2}}{2}\sum_{f} Z_{\textup{1F}}^{f}\int_{q} \, \Tr\!\left[\gamma_{\mu} \+ S_{f}(q) \+\Gamma_{\nu}^{f}(p,q,k) \+ S_{f}(p)\right] 
\end{align}
with the shorthand $\int_{q} \equiv \int \dd^{4}q \+ / \+ (2\pi)^{4}$, a trace over the Dirac part in the gluon DSE, and $\sum_{f}$ represents the sum over the $\Nf=2+1$ quark flavors
$f\in\{\textup{u},\textup{d},\textup{s}\}$ considered in this work. 
The momentum routing is $p=q-k$. Taking the color trace, one obtains the Casimir
$C_{\textup{F}}= (\Nc^{2}-1) \+ / \+ (2\Nc)$ in the quark DSE, where $\Nc=3$ is the number of colors. For the strong coupling constant 
we use $g^{2}=4\pi\alpha_{\textup{s}}$ with $\alpha_{\textup{s}}=0.3$ representing the running coupling 
at a scale fixed by the quenched gluon from the lattice. In the renormalization constant for the quark-gluon vertex
$Z_{\textup{1F}}^{f} = Z_{2}^{f} \tilde{Z}_{1} \+ / \+ \tilde{Z}_{3}$, the ghost-gluon vertex renormalization 
constant $\tilde{Z}_{1}$ is set to one since the corresponding vertex is ultraviolet finite in the 
Landau gauge \cite{Taylor:1971ff} used in this work.

The inverse bare quark propagator at finite quark chemical potential $\mu_{\textup{q}}^{f}$ 
but vanishing temperature $T$ is given by
\begin{align}
S_{f,0}^{-1}(p)=Z_{2}^{f}\left(i\vec{\slashed{p}}+i\tilde{p}_{4}\gamma_{4}+Z_{m}^{f}m_{\textup{R}}^{f}(\xi)\right)\,.
\label{eq:inverse_bare_quark_propagator_finite_mu_only}
\end{align}
$Z_{2}^{f}$ and $Z_{m}^{f}$ are the quark wave function and the quark mass renormalization constants and $m_{\textup{R}}^{f}(\xi)$ is the renormalized current-quark mass of the flavor $f$ at the renormalization point $\xi=\SI{80}{\giga\eV}$. 
Details for the renormalization procedure as well as the numerical set-up used to solve the DSEs can be found in Ref.~\cite{Isserstedt:2019pgx} and references therein. We work with non-zero light-quark chemical potential only, 
i.e., we keep $\muq^\ups=0$. Furthermore, we set the isospin chemical potential to zero, implying
$\muq^\upu=\muq^\upd \equiv \muq^\ell$. In this case, the quark chemical potential is connected to the baryon chemical potential through $\muB=3\muq^\ell$. 

The remaining quantity needed to close the system of equations is the dressed quark-gluon vertex
$\Gamma_{\nu}^{f}(p,q,k)$. In previous works \cite{Fischer:2012vc,Fischer:2014ata,Eichmann:2015kfa}
the construction 
\begin{align}
\label{eq:vertex_ansatz}
\Gamma_{\nu}^{f}(p,q,k) &= \tilde{Z}_{3} \+ \gamma_{\nu} \+ \Gamma(x) \+ \Gamma_{\textup{BC}}^{f}(p,q) \, , \nonumber\\[0.5em]
\Gamma(x) &= \frac{d_{1}}{d_{2}+x\Lambda^{2}}+\frac{x}{1+x}\left(\frac{\beta_{0}\alpha_{\textup{s}}}{4\pi}\ln(1+x)\right)^{2\delta} \, , \nonumber\\[0.5em]
\Gamma_{\textup{BC}}^{f}(p,q) &= (1-\delta_{\nu4}) \+ \frac{A_{f}(p)+A_{f}(q)}{2} \\[0.25em]
& \phantom{=\;} + \delta_{\nu4} \+ \frac{C_{f}(p)+C_{f}(q)}{2} \nonumber
\end{align}
has been used with quark momenta $p$, $q$ and gluon momentum $k$. 
Here $\Gamma(x)$ is an ansatz that captures the well known perturbative running of the vertex at large momenta
and parameterizes the non-perturbative enhancement of the vertex at small momenta known from explicit solutions
of the vertex DSE in vacuum, see, e.g., Refs.~\cite{Braun:2014ata,Williams:2015cvx} and references therein. The 
momentum variable $x$ represents $x= k^{2} / \+ \Lambda^2$ in the quark DSE and 
$x=(p^{2}+q^{2}) \+ / \+ \Lambda^{2}$ in the gluon DSE in order to comply with multiplicative renormalizability.
The leading-order coefficient of the beta function is given by $\beta_{0} = (11\Nc-2\Nf) \+ / \+ 3$. 
The temperature-independent parameters $d_{2}=\SI{0.5}{\giga\eV\squared}$ and $\Lambda=\SI{1.4}{\giga\eV}$ are 
fixed to match the scales in the quenched gluon propagator from the lattice. The parameter $d_{1}$ is the 
effective infrared vertex interaction strength, which will be discussed below. 

$\Gamma_{\textup{BC}}^{f}(p,q)$ together with the tensor structure $\gamma_{\nu}$ represent the leading term 
of the Ball-Chiu vertex construction \cite{Ball:1980ay} satisfying the Abelian Ward-Takahashi identity. 
This term introduces a non-trivial temperature and chemical-potential dependence of the vertex via the quark dressing functions.

In Refs.~\cite{Fischer:2012vc,Fischer:2014ata,Eichmann:2015kfa,Isserstedt:2019pgx} the phase diagram of QCD 
has been calculated in this framework without further approximations, i.e., taking into account the 
chemical-potential dependence of the gluon
via the back-reaction of the quarks in the quark-loop diagrams and using the vertex ansatz \eqref{eq:vertex_ansatz}
consistently in the quark and gluon DSE. In this work, however, due to the numerical complexity induced by 
the additional study of mesons (see Sec.~\ref{sec:meson_bound_states}) we need to introduce two further
approximations. First, we neglect the chemical-potential dependence of the gluon. We have checked explicitly 
that for baryon chemical potentials up to \SI{1}{\giga\eV} this approximation affects the gluon
self-energy
and the gluon screening mass at most on the five-percent level. 
Second, we replace $\Gamma_{\textup{BC}}^{f}(p,q) \rightarrow Z_2^{f}$ in the quark DSE. This is accompanied by an appropriate change of $d_1 \rightarrow d_1^{\+\textup{q}}$. Consequently, this corresponds to a rainbow-ladder approximation 
and greatly facilitates the construction of the Bethe-Salpeter 
kernel\footnote{For details how to construct 
the much more complicated kernel corresponding to the choice of a Ball-Chiu vertex see \cite{Heupel:2014ina}}.
However, in order to maintain contact with previous works 
as much as possible we keep the full construction (\ref{eq:vertex_ansatz}) present in the quark loop
of the gluon DSE. We also take over the vertex strength $d_1=\SI{8.49}{\giga\eV\squared}$ in 
the quark loop of the gluon DSE determined in Ref.~\cite{Isserstedt:2019pgx}.

For the quark masses
$m_{\ell} \equiv m_{\upu}=m_{\upd}$ and $m_{\ups}$, the vertex strength parameter $d_1^{\+\textup{q}}$ is fixed using lattice results for the subtracted quark condensate 
\begin{align}\label{cond1}
	\Delta_{\ell\ups}
	&=
	\expval{\bar{\Psi} \Psi}_\ell
	-
	\frac{m_\ell}{m_\ups} \+ \expval{\bar{\Psi} \Psi}_\ups \, ,\\\label{cond2}
	\expval{\bar{\Psi}\Psi}_f &= -\Nc \+ Z_2^f Z_m^f \+ \SumInt_q \Tr \bigl[ S_f(q) \bigr]
\end{align}
at finite temperature and vanishing chemical potential. The shorthand $\SumInt_q \equiv T\sum_{\omega_n} \int\dd^3 q \+ / \+ (2\pi)^{3}$ includes a sum over discrete Matsubara frequencies $\omega_n = (2n+1)\+\pi T$ ($n \in \mathbb{Z}$) appearing at finite temperature according to $q = (\vec{q},\omega_n)$.
We adapt the vertex strength parameters such that the pseudocritical temperature found on the lattice 
\cite{Borsanyi:2010bp,Bazavov:2011nk,Bellwied:2015rza,Bazavov:2018mes} is reproduced. The light- and strange-quark masses are determined by matching the experimental pion 
and kaon masses in vacuum. The resulting light-to-strange quark mass ratio is
$r_{\ell\ups}=m_\ups \+ / \+ m_\ell = 25.7$. All values for vertex strength and quark masses for the truncation used 
in this work (RL) are given in Tab.~\ref{tab:truncation_parameter} together with the corresponding values 
for the truncation in Eq.~\eqref{eq:vertex_ansatz} (1BC) using $\Gamma_{\textup{BC}}^{f}(p,q)$ in both, the 
quark and gluon DSE employed in previous works.

\begin{table}[t]
\begin{ruledtabular}
	\begin{tabular}{lccccc}
		Sets & $d_{1}^{\+\textup{q}}\;[\textup{GeV}^{2}]$ & $d_{1}\;[\textup{GeV}^{2}]$ & $m_{\ell}\;[\textup{MeV}]$ & $m_{\textup{s}}\;[\textup{MeV}]$ & $r_{\ell\textup{s}}$ \\
		\cmidrule{1-1}
		\cmidrule{2-3}
		\cmidrule{4-5}
		\cmidrule{6-6}
		1BC &  \multicolumn{2}{c}{8.49} & 0.80 & 20.6 & 25.7 \\
		RL  & 12.85 & 8.49 & 1.47 & 37.8 & 25.7 \\
	\end{tabular}
	\caption{\label{tab:truncation_parameter}%
	Vertex strength parameters and quark masses for the truncation set used in previous works \cite{Isserstedt:2019pgx} (1BC; vertex construnction \eqref{eq:vertex_ansatz} in quark and gluon DSEs)
	and the one employed in this work (RL; rainbow-ladder in the quark but full construction in gluon DSE). See the main text for more explanations.}
\end{ruledtabular}
\end{table}
\begin{figure}[t]
 \centering
 \includegraphics[width=0.45\textwidth]{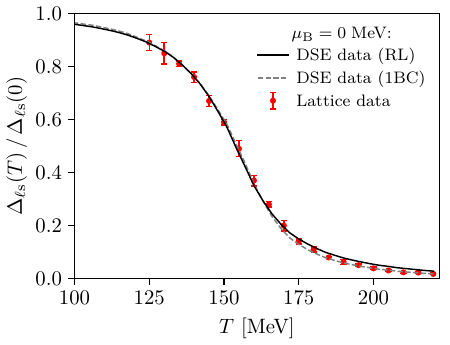}%
 \vspace{-1em}
 \caption{\label{fig:condensate_finite_T_only_fixing_procedure_lattice}%
 Comparison of the vacuum-normalized subtracted quark condensate of the 1BC (dashed gray) \cite{Isserstedt:2019pgx} and RL (solid black) truncated DSE calculation to corresponding continuum-extrapolated lattice results (solid red circles) from Ref.~\cite{Borsanyi:2010bp} at vanishing chemical potential.}
\end{figure}

Our results for the quark condensate determined in both schemes are shown in
Fig.~\ref{fig:condensate_finite_T_only_fixing_procedure_lattice}. For low temperatures both results  
are almost indistinguishable and agree very well with the lattice results of Ref.~\cite{Borsanyi:2010bp}. 
At large temperatures the simplified scheme (RL) produces a slightly too large condensate, which is 
a direct consequence of the somewhat large light-quark mass necessary to reproduce the physical pion 
mass at zero temperature. In contrast, the result for the 1BC truncation taken from 
Ref.~\cite{Isserstedt:2019pgx} has been obtained using a light-quark mass that has been adapted to
match the large-temperature behavior of the condensate to the lattice. In general, in the
high-temperature phase high-quality vertex constructions even beyond the 1BC-level are important.
This is underlined by the study of sensitive observables such as baryon number fluctuations
\cite{Isserstedt:2019pgx}. In this work, however, we are concerned with the zero-temperature case,
where the simpler RL truncation is sufficient to produce meaningful results. 

For the calculation of bound states we need the quark propagator for complex momenta. As discussed
below, quarks in the vacuum Bethe-Salpeter equation are tested at momenta $q_{\pm}= q \pm P/2$, where $q$ 
is space-like and real. The total meson momentum $P$, however, is time-like and given by $P=(\vec{0},iM)$ 
in the rest-frame of the meson with $M$ the mass of the bound state. It is easy to see
that in integrating over $q$, $q_{\pm}^2$ covers a parabola in the complex momentum plane with apex
at real, time-like momenta and symmetry with respect to the real axis. At finite chemical potential the
quark is probed at $q_\pm = q + Q_\pm \+ / \+ 2$ with $Q_\pm = \bigl(\vec{0}, i \+ (2\mu_\textup{q}^f \pm M)\bigr)$, which broadens
the parabola. Inside the parabola, the quark propagator in vacuum as well as for finite chemical potential can therefore be calculated by standard techniques reviewed in Ref.~\cite{Sanchis-Alepuz:2017jjd}.

\subsection{\label{sec:meson_bound_states}Meson bound states}

\begin{figure}[t]
 \centering
 \includegraphics[width=0.45\textwidth]{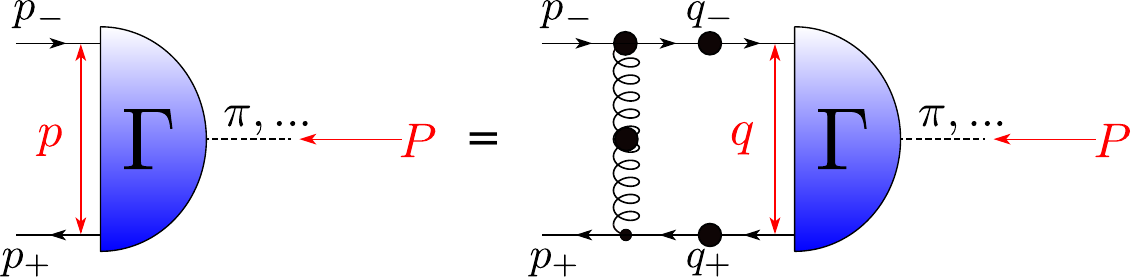}
 \caption{\label{fig:BSE_with_momentum_routing}%
 Graphical representation of the homogeneous Bethe-Salpather equation
 with the corresponding momentum routing.}
\end{figure}

To calculate the wave function and properties of mesonic bound states in the Dyson-Schwinger approach, 
we use the homogeneous Bethe-Salpeter equation (BSE) which is given by
\begin{align}
\Gamma_{X}^{f}(p,P)=-Z_{\textup{1F}}^{f}\+4\pi\alpha_{\textup{s}}\+C_{\textup{F}}&\int_{q}\,\gamma_{\mu}\+S_{f}(q_{+})\+\Gamma_{X}^{f}(q,P) \nonumber \\
&\times S_{f}(q_{-})\+\Gamma_{\nu}^{f}(p,q,k)\+D_{\mu\nu}(k)
\label{eq:homogeneous_Bethe_salpeter_equation}
\end{align}
and displayed in Fig.~\ref{fig:BSE_with_momentum_routing}. The quark momenta $q_{\pm}$ are given 
by $q_{\pm}= q + Q_\pm \+ / \+2$. 
The Bethe-Salpeter amplitude (BSA) $\Gamma_{X}^{f}(p,P)$ of either pseudoscalar or scalar mesons, 
$X \in \{\textup{P},\textup{S}\}$, depends on the relative momenta $p$ between the quark and the antiquark as well 
as on the total meson momentum $P$. The constants $\alpha_{\textup{s}}$, $C_{\textup{F}}$, and 
the quark-gluon vertex renormalization $Z_{\textup{1F}}^{f}$ are the same as in the quark DSE. 
Since the dressing $\Gamma_{\nu}^f(p,q,k)$ of the quark-gluon vertex in the BSE equals the one employed in the quark DSE and consequently is given 
by Eq.~\eqref{eq:vertex_ansatz} together with the replacement $\Gamma_{\textup{BC}}^{f}(p,q) \rightarrow Z_2^{f}$ and the new vertex strength $d_{1}^{\+\textup{q}}$,
it is independent of the two quark momenta and therefore resembles a rainbow-ladder approximation, 
as discussed above. The interaction then satisfies the axial-vector Ward-Takahashi identity, 
and the pion retains its (pseudo-)Goldstone-boson nature.

For the pseudoscalar (P) and scalar (S) mesons in vacuum the Dirac tensor decomposition of the BSAs is given by
\begin{align}
\Gamma_{\textup{P}}(p,P)=\;\gamma_{5}&\left\{E_{\textup{P}}(p,P)-i\slashed{P}F_{\textup{P}}(p,P)\right. \nonumber \\
&\;-\left.i\slashed{p}\,P\cdotp p\,G_{\textup{P}}(p,P)+\left[\slashed{P},\slashed{p}\right]H_{\textup{P}}(p,P)\right\}, \label{eq:BSA_vacuum_pseudoscalar}\\
\Gamma_{\textup{S}}(p,P)=\,\;\openone&\left\{E_{\textup{S}}(p,P)-i\slashed{P}\,P\cdotp p\,F_{\textup{S}}(p,P)\right. \nonumber \\
&\;-\left.i\slashed{p}\,G_{\textup{S}}(p,P)+\left[\slashed{P},\slashed{p}\right]H_{\textup{S}}(p,P)\right\}.
\label{eq:BSA_vacuum_scalar}
\end{align}
The additional flavor dependence is suppressed. In this decomposition, the $\gamma_{5}$-factor of the pseudoscalar mesons accounts for the correct parity transformation. 
The extra factors $P \cdotp p$ at some places ensure that all BSA components $E_X$, $F_X$, $G_X$, and $H_X$ 
with $X \in\{\textup{P},\textup{S}\}$
transform similarly under parity transformation. Furthermore, the signs are chosen such that all BSA components are positive. As explained above, we use the BSE in the vacuum to fix the quark masses that
reproduce the experimental pion and kaon masses. It is well known that the meson masses are mainly
generated by $E_X$, $F_X$ and contributions from $G_X$, $H_X$ are almost negligible. There are, 
however, substantial contributions of $G_X$ to the decay constants.

At finite chemical potential the number of independent tensor structures increases from four to eight:
$\{ \openone, \gamma_5 \} \times \{ \openone, \gamma_{4},\vec{\slashed{P}},\vec{\slashed{p}},
\vec{\slashed{P}}\vec{\slashed{p}},\gamma_4 \, \vec{\slashed{p}},\vec{\slashed{P}}\+\gamma_4, \vec{\slashed{P}} \vec{\slashed{p}} \, \gamma_4 \}$.
Note that the new structure $\gamma_{4}$ arises from both vacuum structures $\slashed{P}$ and $\slashed{p}$. 
Numerically this is very demanding and we only consider the following reduced tensor decomposition:
\begin{align}
\Gamma_{\textup{P}}(p,P)=\;\gamma_{5}&\left\{E_{\textup{P}}(p,P)-i\gamma_{4} \+ I_{\textup{P}}(p,P)\right\}, \\
\Gamma_{\textup{S}}(p,P)=\,\,\;\openone&\left\{E_{\textup{S}}(p,P)-i\gamma_{4} \+ P\cdotp p\+I_{\textup{S}}(p,P)\right\} .
\label{eq:BSA_finite_mu}
\end{align}
In the rest frame of the meson, this choice covers all tensor structures arising from the $F_X$-components 
in the vacuum (since then $\vec{P}=0$) and picks up part of the contributions from $G_X$. Since
meson masses are dominated by $E_X$ and $F_X$, we expect the medium results to be roughly comparable with 
an $EF$-only calculation in the vacuum. 

The BSE components $g_X \in \{E_X,F_X,G_X,H_X,I_X\}$ depend on the total momentum $P$ of the 
bound state, the relative momentum $p$ between the quark and the antiquark, and the angle $z=P\cdotp p \+ / \+ (\lvert p \rvert \lvert P \rvert)$
between these two momenta. Since the momentum direction of the rest 
frame of the meson (which we consider only) and the direction of the medium are aligned we do not
encounter any additional dependencies when we switch on the chemical potential. In vacuum,
the dependence on the angle $z$ turns out to be mild. Therefore an expansion in terms of Chebyshev 
polynomials $T_j$, viz.
\begin{equation}
g_X(P^2,p^2,z) = \sum_j \, g^{\+j}_X(P^2,p^2) \, T_j(z)
\label{eq:chebychev_expansion}
\end{equation}
is appropriate, since it converges rapidly. We will use this expansion also at finite chemical potential
and explore its validity.

The meson BSE is used to extract the meson mass $M_{X}$ and its Bethe-Salpeter amplitude,
see Ref.~\cite{Eichmann:2016yit} for technical details. The amplitude is normalized using the 
Nakanishi method \cite{Nakanishi:1965zza} and the meson decay constant $f_{X}$ in vacuum is calculated via 
\begin{align}
f_{X}P_{\mu}&=Z_{2}^f \+ \Nc\int_{q} \, \Tr \bigl[ \+ j^{X}_{\mu} S_f(q_{+}) \+ \hat{\Gamma}_{X}^f(q,P) \+ S_f(q_{-}) \bigr] \, , \\
j^{X}_{\mu}&=\begin{cases}
             \gamma_{5}\gamma_{\mu} & \textup{for } X=\textup{P}\,, \\
             \gamma_{\mu} & \textup{for } X=\textup{S}\,,
            \end{cases}
\label{eq:meson_decay constant_vacuum}
\end{align}
with the normalized BSA $\hat{\Gamma}_{X}$. Introducing chemical potential, 
the decay constant splits up into two parts \cite{Son:2001ff,Son:2002ci}:
\begin{align}
f_{X}P_{\mu} \, \xrightarrow{\mu_{\textup{B}}\neq0} \, \left[f_{X}^{\textup{s}}T_{\mu\nu}(v)+f_{X}^{\textup{t}}L_{\mu\nu}(v)\right] \? P_{\nu}
\label{eq:decay_constant_splitting_finite_mu}
\end{align}
whereby the spatial $f_{X}^{\textup{s}}$ and temporal $f_{X}^{\textup{t}}$ meson decay constants are transversal and longitudinal to the assigned direction of the medium $v=(\vec{0},1)$.

\section{\label{sec:results}Results}

In the following, we first discuss the quark condensate and the quark dressing functions at 
finite chemical potential and vanishing temperature. Then we study the meson Bethe-Salpeter 
amplitude and the meson properties. 

\subsection{\label{sec:quark_properties_results}Quark properties}
%
\begin{figure}[t]
\centering
\includegraphics[width=0.45\textwidth]{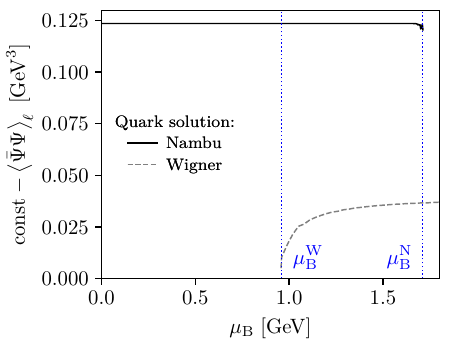}%
\vspace{-1em}
\caption{\label{fig:quark_condensate_finite_mu}%
Shifted light-quark condensate at finite chemical potential for the chirally-broken Nambu 
(solid black) and chirally-restored Wigner (dashed gray) solution. The boundaries for the
appearance/disappearance of these solutions are denoted by $\mu_{\textup{B}}^{\textup{N},\textup{W}}$ and indicated by vertical dotted lines. 
}
\end{figure}
\begin{figure*}[t]
\centering
\includegraphics[width=0.45\textwidth]{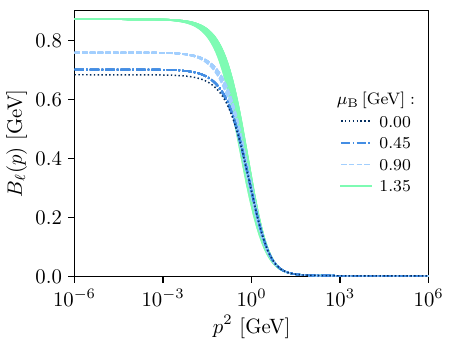}
\hfill
\includegraphics[width=0.45\textwidth]{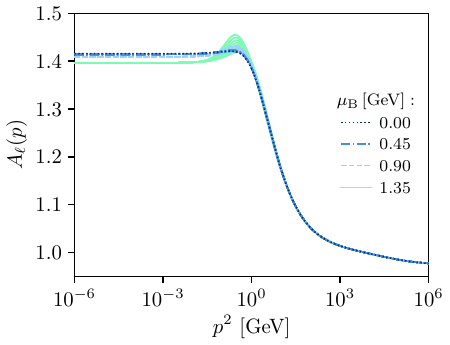}%
\vspace*{-1em}
\caption{\label{fig:quark_dressing_functions_finite_mu}%
Scalar $B_\ell$ (left) and first vector $A_\ell$ (right) light-quark dressing function for different baryon chemical potentials $\mu_{\textup{B}}$ against real and space-like four-momenta $p^2$. 
All quark dressing functions correspond to the Nambu solution. The spread of the dressing functions corresponds to their angular dependence due to the assigned direction of the medium.}
\vspace*{1em}
\includegraphics[width=0.45\textwidth]{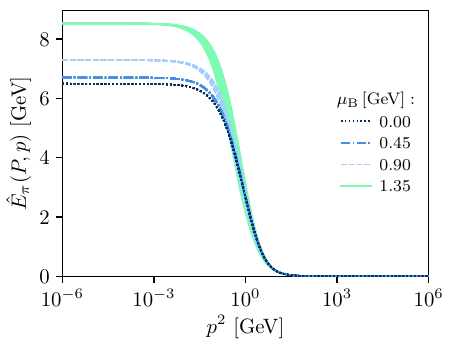}
\hfill
\includegraphics[width=0.45\textwidth]{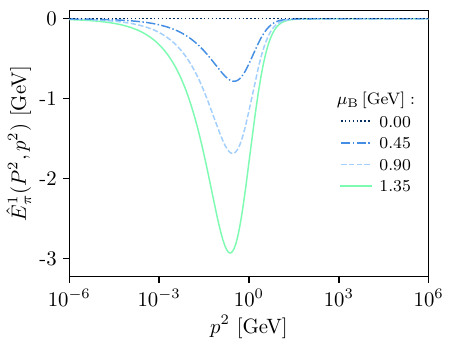}%
\vspace*{-1em}
\caption{\label{fig:BSA_finite_mu}%
First normalized on-shell pion BSA component $\hat{E}_{\pi}$ (left) and corresponding first Chebychev coefficient $\hat{E}_{\pi}^{1}$ of it (right) against 
the relative momentum $p^{2}$ between the quarks for various baryon chemical potentials $\mu_{\textup{B}}$ below the first-order phase transition. In the left diagram, the spread of the amplitude corresponds to its dependence on the angle between $P$ and $p$.}
\vspace*{1em}
\centering
\includegraphics[width=0.45\textwidth]{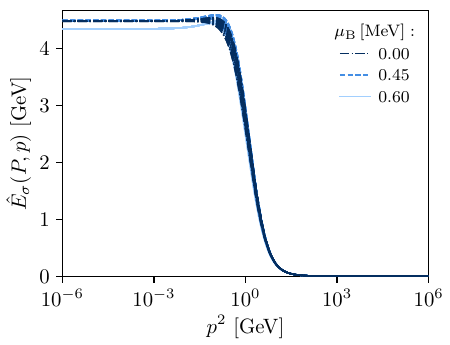}%
\vspace*{-1em}
\caption{\label{fig:B_over_fpi_sigma_BSA}%
First normalized on-shell sigma BSA component $\hat{E}_{\sigma}$ against the relative momentum $p^{2}$ between the quarks for various chemical potentials $\mu_{\textup{B}}$ below the first-order phase transition. Again, the spread corresponds to its dependence of the angle between $P$ and $p$.}
\end{figure*}

In Fig.~\ref{fig:quark_condensate_finite_mu} we show our results for the light-quark condensate
$\expval{\bar{\Psi}\Psi}_{\ell}$ (shifted by a constant) against the baryon chemical 
potential $\mu_{\textup{B}}$. The condensate plotted is determined from two types of stable 
solutions of the quark DSE: one is the chirally broken (Nambu) and the other the chirally 
symmetric (Wigner) solution. The appearance/disappearance of these solutions mark three
different regions in the plot. Below $\mu_{\textup{B}}^{\textup{W}}$ only the Nambu solution 
is found in the numerical iteration\footnote{The Wigner solution is also present, but is not 
iteratively attractive and thermodynamically not favoured, see, e.g.,
Refs.~\cite{Williams:2006vva,Chang:2006bm,Fischer:2008sp} for further details and explanations on the
appearance of different solutions in the quark DSE.}, whereas above $\mu_{\textup{B}}^{\textup{N}}$ 
the Nambu solution disappears and only the Wigner solution remains. 
In between $\mu_{\textup{B}}^{\textup{W}}$ and $\mu_{\textup{B}}^{\textup{N}}$, we find both solutions. The boundaries of this mixed region are given by 
\begin{equation}
\begin{aligned}
\mu_{\textup{B}}^{\textup{W}}=&\;\SI{0.956}{\giga\eV}, \\
\mu_{\textup{B}}^{\textup{N}}=&\;\SI{1.715}{\giga\eV}\,.
\end{aligned}
\end{equation}
The physical phase transition between the Nambu and the Wigner solution happens in the mixed region
and has to be determined from thermodynamical considerations, which are outside the scope of the 
present work. 

The quark condensate of the Nambu solution remains constant for chemical potentials below
$\mu_{\textup{B}}^{\textup{W}}$ and also in the majority of the mixed region between 
$\mu_{\textup{B}}^{\textup{W}}$ and $\mu_{\textup{B}}^{\textup{N}}$. Since
$\mu_{\textup{B}}^{\textup{W}}$ is larger than the nucleon mass $m_\textup{N}$, the condensate therefore 
satisfies the Silver-Blaze property, although strictly speaking it is not an observable quantity. 
The precision is astonishing: up to $\mu_{\textup{B}}=m_\textup{N}$ the shifted quark condensate only 
increases by $0.001\,\%$.
 
The quark condensate for the Wigner solution increases for lower chemical potentials and settles 
for higher ones. However, it is not clear to what extent this calculation can be trusted. As 
explained above, the introduction of finite chemical potential induces a shift in the momenta of
the quark into the complex momentum plane. As analyzed in detail in Ref.~\cite{Fischer:2008sp}, the
Wigner solution in the vacuum features a pair of complex-conjugate poles close to the real axis at very low quark
masses. For the large chemical potential considered here, the integration path in the quark DSE
is shifted beyond these poles, resulting in residues which have not yet been taken into account
properly. Since the main emphasis of this paper is on the Nambu solution, this is left for future 
work. 

The quark dressing functions $A_\ell$ and $B_\ell$ for the Nambu solution for light quarks are shown in
Fig.~\ref{fig:quark_dressing_functions_finite_mu}. For momenta much larger than the scale of the 
chemical potential the dressing functions are almost independent of chemical potential. At small 
momenta the scalar quark dressing function $B_\ell$ from the Nambu solution increases drastically 
with chemical potential. The Wigner solution (not show in the plot) is much smaller in magnitude and has 
almost no dynamical contribution. The vector quark dressing function $A_\ell$ is much less 
dependent on chemical potential, although also here effects can be seen. The dressing function 
$C_\ell$ is identical to $A_\ell$ for the Nambu solution. 
Due to this degeneracy the number of independent Dirac tensor structures  
is the same as in the vacuum. This property of the quark has already been shown analytically for 
the case of the rainbow-ladder truncation in Ref.~\cite{Zong:2005mm}. It is therefore no surprise 
that it also shows up in our numerical calculation. Whether it still holds when chemical potential 
effects in the gluon and in the quark-gluon vertex are considered remains to be investigated. 

\subsection{\label{sec:bethe_salpeter_amplitudes_results}Bethe-Salpeter amplitues}

The first normalized component of the pion BSA $E_\pi$ is shown in the left diagram of Fig.~\ref{fig:BSA_finite_mu}.
Its dependence on chemical potential is similar to the one of the scalar quark dressing function. 
This is readily explained by the exact relation $E_\pi=B / f_\pi$ in the chiral limit at zero chemical potential
\cite{Maris:1997hd}, which is approximately true also for finite quark masses and is expected to 
persist also at finite chemical potential. At small momenta we observe a strong increase of $E_\pi$ 
with chemical potential.

Of all the coefficients $E^j_\pi(P^2,p^2)$ from the Chebychev expansion \eqref{eq:chebychev_expansion}, the zeroth is the only one with a non-vanishing 
contribution at very low momenta. At small chemical potential it gives by far the largest contribution
to the full amplitude shown on the left side of Fig.~\ref{fig:BSA_finite_mu}. All higher coefficients 
mainly contribute to the chemical-potential dependence of the BSA in the mid-momentum region. 
Their importance increases with chemical potential as can be seen for the example of $E^1_\pi$
in the right diagram of Fig.~\ref{fig:BSA_finite_mu}. 

A further interesting aspect is charge-conjugation symmetry. At zero chemical potential, the neutral 
pion is an invariant state under charge conjugation. As detailed in the appendix,  
this causes all odd Chebychev coefficients to vanish. Consequently, $E^1_\pi(\muB=0) = 0$.
As soon as chemical potential is switched on, however, this property disappears and the odd coefficients
begin to contribute. In fact for large chemical potentials $E_\pi^1$ becomes even more significant 
than $E^2_\pi$.

The second pion BSA component $I_\pi$ (without plot) behaves in the same fashion as the first one but shows an 
even stronger chemical-potential dependence in the small and mid momentum regime. 
Since the $I_\pi$-BSA component gets contributions from the 
$F_\pi$- and $G_\pi$-BSA components, we cannot compare it with the vacuum. The first BSA component
of the sigma meson on the other hand does not show a strong chemical potential dependency, as seen in
Fig.~\ref{fig:B_over_fpi_sigma_BSA}, whereas the second one is similar to the one of the pion.

\subsection{\label{sec:meson_properties_results}Meson properties}

\begin{figure}
\centering
\includegraphics[width=0.45\textwidth]{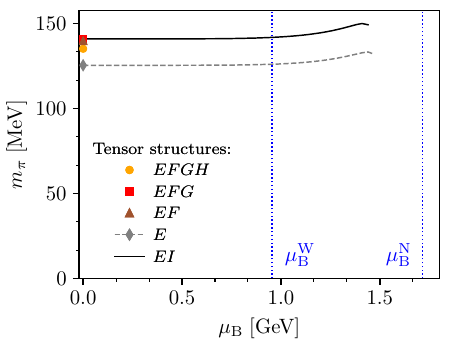}\\
\includegraphics[width=0.45\textwidth]{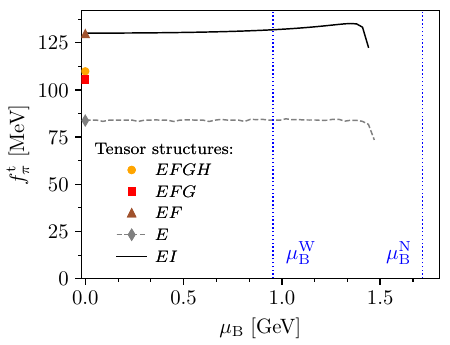}%
\vspace*{-1em}
\caption{\label{fig:pion_properties_finite_mu}%
Pion mass (upper) and temporal pion decay constant (lower) against the baryon 
chemical potential for different combinations of tensor structures used in the BSE calculation. 
The colored dots represent the vacuum BSE calculations. All results are calculated with the
chirally-broken Nambu solution of the quark DSE.}
\end{figure}

In this section, we use the results from the last two sections to study the meson properties of the 
pion and sigma meson. Since we consider the meson to be in the rest frame, we are only able to study the temporal decay constant
and not the spatial one.

In Fig.~\ref{fig:pion_properties_finite_mu} we display the pion mass and temporal pion decay constant 
in vacuum and at finite chemical potential. In the vacuum, we denote the results for different levels
of approximation in the BSA as colored symbols. As explained above, due to numerical complexity we only 
use two tensor components at finite chemical potential. As can be seen from the plots, this choice 
already leads to quantitatively results for the pion mass, whereas the results for the decay constant
remain qualitative. 

We only
determined the pion properties from the Nambu solution. In principle one could also use the Wigner
solution in the BSE, however we refrain from doing so for two reasons: (i) there are technical problems
related to the analytic structure of the quark propagator for this solution; (ii) for chemical potentials
larger than the one of the chiral phase transition, $\muB > \mu_{\textup{B}}^\textup{1st}$,
one expects to encounter a color-superconducting phase, which requires a substantial extension of 
the framework to the Nambu-Gorkov formalism studied elsewhere \cite{Muller:2013pya,Muller:2016fdr}.

Within numerical precision, the pion mass and temporal decay constant of both approximations for the 
BSA ($E$ and $EI$) remain constant beyond the baryon chemical potential of a nucleon and therefore
satisfy the Silver-Blaze property. Deviations are smaller than $1\,\%$ for the pion mass and $2\,\%$
for the temporal decay constant. Inside the mixed region, the variations become larger until at some 
point the numerical stability of the solutions decreases drastically and we are no longer able to
obtain solutions beyond that point. It is not clear whether this scale is associated with the appearance
of technical problems or with the first-order phase transition. This needs to be explored in future 
work. It is also interesting to see that the inclusion of tensor structures beyond $E$ leads to 
numerically more stable results than using $E$ alone. 

\begin{figure}
\centering
\includegraphics[width=0.45\textwidth]{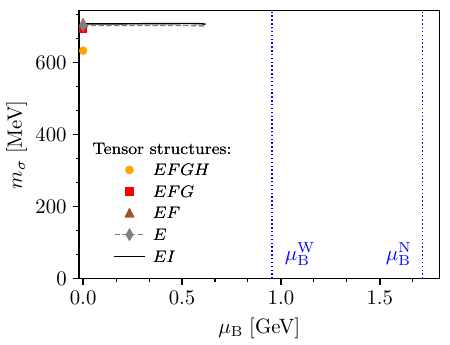}%
\vspace*{-1em}
\caption{\label{fig:sigma_properties_finite_mu}%
Sigma mass against the baryon chemical potential for different numbers of used tensor structures in the BSE calculation, as described in Fig.~\ref{fig:pion_properties_finite_mu}. Again, all results are obtained using the chirally-broken Nambu solution of the quark DSE.}
\end{figure}
Our results for the mass of the sigma meson are shown in Fig.~\ref{fig:sigma_properties_finite_mu}.
Here, the technical problems associated with the analytic structure of the quark propagator become 
apparent already for smaller chemical potentials and we can determine the sigma mass only 
up to approximately $\muB=\SI{600}{\mega\eV}$. In the plot, the results for the two
BSA combinations, $E$ and $EI$, are almost similar and can hardly be distinguished. We also see
that the Silver-Blaze property is satisfied within the range of chemical potential that is accessible 
in our calculation. 

We wish to emphasize that the observation of the Silver-Blaze property for meson observables 
is not new, but has been observed before at least approximately in previous works using
functional methods \cite{Maris:1997eg,Bender:1997jf,Detmold:2002,Jiang:2008rb}. 
The new element in our calculation as compared to the latter works is the explicit numerical 
solution of the quark DSE in the complex momentum plane (taking into account the full momentum 
dependence of the gluon) and the subsequent solution of both, the meson BSE and the equation 
for the temporal decay constant. In previous works, apart from Ref.~\cite{Maris:1997eg}, 
only the approximation $E_\pi=B/f_\pi$ for the meson BSA has been studied. A contact interaction 
between quarks has been employed in Ref.~\cite{Maris:1997eg} which neglects all momentum dependence 
of the gluon. Moreover, Ref.~\cite{Detmold:2002} is the only work in this context that takes the splitting 
of the decay constant into temporal and spatial parts into account. Finally, the effects of 
tensor structures beyond $E$ in the meson BSA has not been studied so far at finite chemical 
potential. In general it is very satisfying to see that the Silver-Blaze property in meson observables 
can be maintained also in elaborated truncation schemes that path the way towards quantitative 
results. 

\section{\label{sec:summary}Summary and conclusions}

In this work we have studied the chemical-potential dependence of the masses and decay constants 
of the pion and the sigma meson in the functional framework of DSEs. To this end we used an elaborate and 
well-studied truncation scheme for the underlying Dyson-Schwinger equations for quarks and gluons 
which incorporates lattice input into functional methods. We have solved these equations for 
complex momenta thereby providing input into Bethe-Salpeter equations (BSEs) that describe the properties
of bound states of quarks and antiquarks. We have solved these equations, and the corresponding ones
for the pion decay constant for chemical potentials up to and inside the coexistence region
of the the first-order phase transition. We traced the existence of the pion as a (pseudo-)Goldstone boson
well into the mixed region of coexistence of the chirally broken and the chirally symmetric phase.
For larger chemical potentials we were not able to identify a solution of the BSEs. Almost
up to this scale, the pion mass and decay constant remain constant under change of the
chemical potential. In the region where our calculations were technically possible, we find the same
behavior for the mass of the sigma meson. This Silver-Blaze property is a non-trivial consequence 
of substantial changes in the quark dressing functions and the meson Bethe-Salpeter amplitudes that 
conspire with each other to cancel out in observable quantities.


\begin{acknowledgments}
    We thank Richard Williams for fruitful discussions.
	This work has been supported by the Helmholtz Graduate School for Hadron and Ion Research
	(HGS-HIRe) for FAIR, the GSI Helmholtzzentrum f\"{u}r Schwerionenforschung, 
	the Helmholtz International Center for FAIR within the LOEWE program of the
	State of Hesse, and the BMBF under contract No.~05P18RGFCA. 
\end{acknowledgments}

\appendix*

\section{Chebyshev expansion and charge-conjugation parity}

As mentioned before the BSE is solved using a Chebychev expansion of the BSA components. We use this Chebychev expansion (Eq.~\eqref{eq:chebychev_expansion}) to study the charge conjugation invariance (C-parity invariance) of the scalar (S) and pseudoscalar (P) mesons. For this purpose, we define the charge conjugated BSA as follows:
\begin{align}
\bar{\Gamma}_{X}(p,P)=\left[C\Gamma_{X}(-p,P)C^{-1}\right]^{\top}=\eta_{X}\Gamma_{X}(p,P)
\label{eq:charge_conjugation_c_parity}
\end{align}
with the `eigenvalue' $\eta_{X}=+1$ for S ($J^{PC}=0^{++}$) and P ($J^{PC}=0^{-+}$) mesons. In this equation, $C=\gamma_{2}\gamma_{4}$ represents the charge conjugation matrix and `$\top$' denotes matrix transposition. Considering the tensor decomposition in Eqs.~\eqref{eq:BSA_vacuum_pseudoscalar} and \eqref{eq:BSA_vacuum_scalar}, the Dirac tensor structures $T_{g_X}^{X}$ transform according to
\begin{align}
\bar{T}_{g_X}^{X}(p,P)=\left[CT_{g_X}^{X}(-p,P)C^{-1}\right]^{\top}=T_{g_X}^{X}(p,P)
\end{align}
with $g_X\in\{E_X,F_X,G_X,H_X\}$. The charge-parity invariance manifests itself in the same symmetry of all BSA components for P and S:
\begin{align}
\bar{g}_X(p^{2},P^{2},z) &= g_X(p^{2},P^{2},-z) \nonumber \\[0.5em]
&= \sum_j \, (-1)^{j} \, g_X^{2j}(p^{2},P^{2}) \, T_{2j}(-z) \\
& \phantom{=\;} + i \, \sum_j \, (-1)^{j} \, g_X^{2j+1}(p^{2},P^{2}) \, T_{2j+1}(-z) \, . \nonumber 
\label{eq:C_parity_inariance_chebychev_coefficients}
\\
&\stackrel{!}{=} g_X(p^2,P^2,z) \nonumber
\end{align}
Since $T_{2j}(z) = T_{2j}(-z)$ but $T_{2j+1}(z) = -T_{2j+1}(-z)$ for all $j$, the odd Chebyshev coefficients have to vanish. To remember, charge conjugation exchanges particles with their corresponding antiparticles and vice versa. Due to the energy offset which is introduced by the chemical potential, antiparticles and particles are no longer energetically degenerated.

\bibliography{quarks_mesons_finite_mu}

\begin{thebibliography}{53}%
\makeatletter
\providecommand \@ifxundefined [1]{%
 \@ifx{#1\undefined}
}%
\providecommand \@ifnum [1]{%
 \ifnum #1\expandafter \@firstoftwo
 \else \expandafter \@secondoftwo
 \fi
}%
\providecommand \@ifx [1]{%
 \ifx #1\expandafter \@firstoftwo
 \else \expandafter \@secondoftwo
 \fi
}%
\providecommand \natexlab [1]{#1}%
\providecommand \enquote  [1]{``#1''}%
\providecommand \bibnamefont  [1]{#1}%
\providecommand \bibfnamefont [1]{#1}%
\providecommand \citenamefont [1]{#1}%
\providecommand \href@noop [0]{\@secondoftwo}%
\providecommand \href [0]{\begingroup \@sanitize@url \@href}%
\providecommand \@href[1]{\@@startlink{#1}\@@href}%
\providecommand \@@href[1]{\endgroup#1\@@endlink}%
\providecommand \@sanitize@url [0]{\catcode `\\12\catcode `\$12\catcode
  `\&12\catcode `\#12\catcode `\^12\catcode `\_12\catcode `\%12\relax}%
\providecommand \@@startlink[1]{}%
\providecommand \@@endlink[0]{}%
\providecommand \url  [0]{\begingroup\@sanitize@url \@url }%
\providecommand \@url [1]{\endgroup\@href {#1}{\urlprefix }}%
\providecommand \urlprefix  [0]{URL }%
\providecommand \Eprint [0]{\href }%
\providecommand \doibase [0]{https://doi.org/}%
\providecommand \selectlanguage [0]{\@gobble}%
\providecommand \bibinfo  [0]{\@secondoftwo}%
\providecommand \bibfield  [0]{\@secondoftwo}%
\providecommand \translation [1]{[#1]}%
\providecommand \BibitemOpen [0]{}%
\providecommand \bibitemStop [0]{}%
\providecommand \bibitemNoStop [0]{.\EOS\space}%
\providecommand \EOS [0]{\spacefactor3000\relax}%
\providecommand \BibitemShut  [1]{\csname bibitem#1\endcsname}%
\let\auto@bib@innerbib\@empty
\bibitem [{\citenamefont {Bors{\'a}nyi}\ \emph {et~al.}(2010)\citenamefont
  {Bors{\'a}nyi} \emph {et~al.}}]{Borsanyi:2010bp}%
  \BibitemOpen
  \bibfield  {author} {\bibinfo {author} {\bibfnamefont {S.}~\bibnamefont
  {Bors{\'a}nyi}} \emph {et~al.},\ }\href
  {https://doi.org/10.1007/JHEP09(2010)073} {\bibfield  {journal} {\bibinfo
  {journal} {J. High Energy Phys.}\ }\textbf {\bibinfo {volume} {2010}},\
  \bibinfo {pages} {73}},\ \Eprint {https://arxiv.org/abs/1005.3508}
  {arXiv:1005.3508 [hep-ph]} \BibitemShut {NoStop}%
\bibitem [{\citenamefont {Bazavov}\ \emph {et~al.}(2012)\citenamefont {Bazavov}
  \emph {et~al.}}]{Bazavov:2011nk}%
  \BibitemOpen
  \bibfield  {author} {\bibinfo {author} {\bibfnamefont {A.}~\bibnamefont
  {Bazavov}} \emph {et~al.},\ }\href
  {https://doi.org/10.1103/PhysRevD.85.054503} {\bibfield  {journal} {\bibinfo
  {journal} {Phys. Rev. D}\ }\textbf {\bibinfo {volume} {85}},\ \bibinfo
  {pages} {054503} (\bibinfo {year} {2012})},\ \Eprint
  {https://arxiv.org/abs/1111.1710} {arXiv:1111.1710 [hep-lat]} \BibitemShut
  {NoStop}%
\bibitem [{\citenamefont {Bhattacharya}\ \emph {et~al.}(2014)\citenamefont
  {Bhattacharya} \emph {et~al.}}]{Bhattacharya:2014ara}%
  \BibitemOpen
  \bibfield  {author} {\bibinfo {author} {\bibfnamefont {T.}~\bibnamefont
  {Bhattacharya}} \emph {et~al.},\ }\href
  {https://doi.org/10.1103/PhysRevLett.113.082001} {\bibfield  {journal}
  {\bibinfo  {journal} {Phys. Rev. Lett.}\ }\textbf {\bibinfo {volume} {113}},\
  \bibinfo {pages} {082001} (\bibinfo {year} {2014})},\ \Eprint
  {https://arxiv.org/abs/1402.5175} {arXiv:1402.5175 [hep-lat]} \BibitemShut
  {NoStop}%
\bibitem [{\citenamefont {Bazavov}\ \emph {et~al.}(2014)\citenamefont {Bazavov}
  \emph {et~al.}}]{Bazavov:2014pvz}%
  \BibitemOpen
  \bibfield  {author} {\bibinfo {author} {\bibfnamefont {A.}~\bibnamefont
  {Bazavov}} \emph {et~al.},\ }\href
  {https://doi.org/10.1103/PhysRevD.90.094503} {\bibfield  {journal} {\bibinfo
  {journal} {Phys. Rev. D}\ }\textbf {\bibinfo {volume} {90}},\ \bibinfo
  {pages} {094503} (\bibinfo {year} {2014})},\ \Eprint
  {https://arxiv.org/abs/1407.6387} {arXiv:1407.6387 [hep-lat]} \BibitemShut
  {NoStop}%
\bibitem [{\citenamefont {Drews}\ and\ \citenamefont
  {Weise}(2017)}]{Drews:2016wpi}%
  \BibitemOpen
  \bibfield  {author} {\bibinfo {author} {\bibfnamefont {M.}~\bibnamefont
  {Drews}}\ and\ \bibinfo {author} {\bibfnamefont {W.}~\bibnamefont {Weise}},\
  }\href {https://doi.org/10.1016/j.ppnp.2016.10.002} {\bibfield  {journal}
  {\bibinfo  {journal} {Prog. Part. Nucl. Phys.}\ }\textbf {\bibinfo {volume}
  {93}},\ \bibinfo {pages} {69} (\bibinfo {year} {2017})},\ \Eprint
  {https://arxiv.org/abs/1610.07568} {arXiv:1610.07568 [nucl-th]} \BibitemShut
  {NoStop}%
\bibitem [{\citenamefont {Fukushima}\ and\ \citenamefont
  {Skokov}(2017)}]{Fukushima:2017csk}%
  \BibitemOpen
  \bibfield  {author} {\bibinfo {author} {\bibfnamefont {K.}~\bibnamefont
  {Fukushima}}\ and\ \bibinfo {author} {\bibfnamefont {V.}~\bibnamefont
  {Skokov}},\ }\href {https://doi.org/10.1016/j.ppnp.2017.05.002} {\bibfield
  {journal} {\bibinfo  {journal} {Prog. Part. Nucl. Phys.}\ }\textbf {\bibinfo
  {volume} {96}},\ \bibinfo {pages} {154} (\bibinfo {year} {2017})},\ \Eprint
  {https://arxiv.org/abs/1705.00718} {arXiv:1705.00718 [hep-ph]} \BibitemShut
  {NoStop}%
\bibitem [{\citenamefont {Fischer}(2019)}]{Fischer:2018sdj}%
  \BibitemOpen
  \bibfield  {author} {\bibinfo {author} {\bibfnamefont {C.~S.}\ \bibnamefont
  {Fischer}},\ }\href {https://doi.org/10.1016/j.ppnp.2019.01.002} {\bibfield
  {journal} {\bibinfo  {journal} {Prog. Part. Nucl. Phys.}\ }\textbf {\bibinfo
  {volume} {105}},\ \bibinfo {pages} {1} (\bibinfo {year} {2019})},\ \Eprint
  {https://arxiv.org/abs/1810.12938} {arXiv:1810.12938 [hep-ph]} \BibitemShut
  {NoStop}%
\bibitem [{\citenamefont {Cohen}\ \emph {et~al.}(1992)\citenamefont {Cohen},
  \citenamefont {Furnstahl},\ and\ \citenamefont {Griegel}}]{Cohen:1991nk}%
  \BibitemOpen
  \bibfield  {author} {\bibinfo {author} {\bibfnamefont {T.~D.}\ \bibnamefont
  {Cohen}}, \bibinfo {author} {\bibfnamefont {R.~J.}\ \bibnamefont
  {Furnstahl}},\ and\ \bibinfo {author} {\bibfnamefont {D.~K.}\ \bibnamefont
  {Griegel}},\ }\href {https://doi.org/10.1103/PhysRevC.45.1881} {\bibfield
  {journal} {\bibinfo  {journal} {Phys. Rev. C}\ }\textbf {\bibinfo {volume}
  {45}},\ \bibinfo {pages} {1881} (\bibinfo {year} {1992})}\BibitemShut
  {NoStop}%
\bibitem [{\citenamefont {Cohen}(2004)}]{Cohen:2004qp}%
  \BibitemOpen
  \bibfield  {author} {\bibinfo {author} {\bibfnamefont {T.~D.}\ \bibnamefont
  {Cohen}},\ }\href@noop {} {}\bibinfo {howpublished}
  {\href{http://arxiv.org/abs/hep-ph/0405043}{arXiv:hep-ph/0405043}} (\bibinfo
  {year} {2004})\BibitemShut {NoStop}%
\bibitem [{\citenamefont {Fromm}\ \emph {et~al.}(2013)\citenamefont {Fromm},
  \citenamefont {Langelage}, \citenamefont {Lottini}, \citenamefont {Neuman},\
  and\ \citenamefont {Philipsen}}]{Fromm:2012eb}%
  \BibitemOpen
  \bibfield  {author} {\bibinfo {author} {\bibfnamefont {M.}~\bibnamefont
  {Fromm}}, \bibinfo {author} {\bibfnamefont {J.}~\bibnamefont {Langelage}},
  \bibinfo {author} {\bibfnamefont {S.}~\bibnamefont {Lottini}}, \bibinfo
  {author} {\bibfnamefont {M.}~\bibnamefont {Neuman}},\ and\ \bibinfo {author}
  {\bibfnamefont {O.}~\bibnamefont {Philipsen}},\ }\href
  {https://doi.org/10.1103/PhysRevLett.110.122001} {\bibfield  {journal}
  {\bibinfo  {journal} {Phys. Rev. Lett.}\ }\textbf {\bibinfo {volume} {110}},\
  \bibinfo {pages} {122001} (\bibinfo {year} {2013})},\ \Eprint
  {https://arxiv.org/abs/1207.3005} {arXiv:1207.3005 [hep-lat]} \BibitemShut
  {NoStop}%
\bibitem [{\citenamefont {Rapp}\ and\ \citenamefont
  {Wambach}(2000)}]{Rapp:1999ej}%
  \BibitemOpen
  \bibfield  {author} {\bibinfo {author} {\bibfnamefont {R.}~\bibnamefont
  {Rapp}}\ and\ \bibinfo {author} {\bibfnamefont {J.}~\bibnamefont {Wambach}},\
  }\href {https://doi.org/10.1007/0-306-47101-9_1} {\bibfield  {journal}
  {\bibinfo  {journal} {Adv. Nucl. Phys.}\ }\textbf {\bibinfo {volume} {25}},\
  \bibinfo {pages} {1} (\bibinfo {year} {2000})},\ \Eprint
  {https://arxiv.org/abs/hep-ph/9909229} {arXiv:hep-ph/9909229 [hep-ph]}
  \BibitemShut {NoStop}%
\bibitem [{\citenamefont {Leupold}\ \emph {et~al.}(2010)\citenamefont
  {Leupold}, \citenamefont {Metag},\ and\ \citenamefont
  {Mosel}}]{Leupold:2009kz}%
  \BibitemOpen
  \bibfield  {author} {\bibinfo {author} {\bibfnamefont {S.}~\bibnamefont
  {Leupold}}, \bibinfo {author} {\bibfnamefont {V.}~\bibnamefont {Metag}},\
  and\ \bibinfo {author} {\bibfnamefont {U.}~\bibnamefont {Mosel}},\ }\href
  {https://doi.org/10.1142/S0218301310014728} {\bibfield  {journal} {\bibinfo
  {journal} {Int. J. Mod. Phys. E}\ }\textbf {\bibinfo {volume} {19}},\
  \bibinfo {pages} {147} (\bibinfo {year} {2010})},\ \Eprint
  {https://arxiv.org/abs/0907.2388} {arXiv:0907.2388 [nucl-th]} \BibitemShut
  {NoStop}%
\bibitem [{\citenamefont {Skokov}\ \emph {et~al.}(2010)\citenamefont {Skokov},
  \citenamefont {Stokic}, \citenamefont {Friman},\ and\ \citenamefont
  {Redlich}}]{Skokov:2010wb}%
  \BibitemOpen
  \bibfield  {author} {\bibinfo {author} {\bibfnamefont {V.}~\bibnamefont
  {Skokov}}, \bibinfo {author} {\bibfnamefont {B.}~\bibnamefont {Stokic}},
  \bibinfo {author} {\bibfnamefont {B.}~\bibnamefont {Friman}},\ and\ \bibinfo
  {author} {\bibfnamefont {K.}~\bibnamefont {Redlich}},\ }\href
  {https://doi.org/10.1103/PhysRevC.82.015206} {\bibfield  {journal} {\bibinfo
  {journal} {Phys. Rev. C}\ }\textbf {\bibinfo {volume} {82}},\ \bibinfo
  {pages} {015206} (\bibinfo {year} {2010})},\ \Eprint
  {https://arxiv.org/abs/1004.2665} {arXiv:1004.2665 [hep-ph]} \BibitemShut
  {NoStop}%
\bibitem [{\citenamefont {Herbst}\ \emph {et~al.}(2011)\citenamefont {Herbst},
  \citenamefont {Pawlowski},\ and\ \citenamefont {Schaefer}}]{Herbst:2010rf}%
  \BibitemOpen
  \bibfield  {author} {\bibinfo {author} {\bibfnamefont {T.~K.}\ \bibnamefont
  {Herbst}}, \bibinfo {author} {\bibfnamefont {J.~M.}\ \bibnamefont
  {Pawlowski}},\ and\ \bibinfo {author} {\bibfnamefont {B.-J.}\ \bibnamefont
  {Schaefer}},\ }\href {https://doi.org/10.1016/j.physletb.2010.12.003}
  {\bibfield  {journal} {\bibinfo  {journal} {Phys. Lett. B}\ }\textbf
  {\bibinfo {volume} {696}},\ \bibinfo {pages} {58} (\bibinfo {year} {2011})},\
  \Eprint {https://arxiv.org/abs/1008.0081} {arXiv:1008.0081 [hep-ph]}
  \BibitemShut {NoStop}%
\bibitem [{\citenamefont {Herbst}\ \emph {et~al.}(2013)\citenamefont {Herbst},
  \citenamefont {Pawlowski},\ and\ \citenamefont {Schaefer}}]{Herbst:2013ail}%
  \BibitemOpen
  \bibfield  {author} {\bibinfo {author} {\bibfnamefont {T.~K.}\ \bibnamefont
  {Herbst}}, \bibinfo {author} {\bibfnamefont {J.~M.}\ \bibnamefont
  {Pawlowski}},\ and\ \bibinfo {author} {\bibfnamefont {B.-J.}\ \bibnamefont
  {Schaefer}},\ }\href {https://doi.org/10.1103/PhysRevD.88.014007} {\bibfield
  {journal} {\bibinfo  {journal} {Phys. Rev. D}\ }\textbf {\bibinfo {volume}
  {88}},\ \bibinfo {pages} {014007} (\bibinfo {year} {2013})},\ \Eprint
  {https://arxiv.org/abs/1302.1426} {arXiv:1302.1426 [hep-ph]} \BibitemShut
  {NoStop}%
\bibitem [{\citenamefont {Tripolt}\ \emph {et~al.}(2018)\citenamefont
  {Tripolt}, \citenamefont {Schaefer}, \citenamefont {von Smekal},\ and\
  \citenamefont {Wambach}}]{Tripolt:2017zgc}%
  \BibitemOpen
  \bibfield  {author} {\bibinfo {author} {\bibfnamefont {R.-A.}\ \bibnamefont
  {Tripolt}}, \bibinfo {author} {\bibfnamefont {B.-J.}\ \bibnamefont
  {Schaefer}}, \bibinfo {author} {\bibfnamefont {L.}~\bibnamefont {von
  Smekal}},\ and\ \bibinfo {author} {\bibfnamefont {J.}~\bibnamefont
  {Wambach}},\ }\href {https://doi.org/10.1103/PhysRevD.97.034022} {\bibfield
  {journal} {\bibinfo  {journal} {Phys. Rev. D}\ }\textbf {\bibinfo {volume}
  {97}},\ \bibinfo {pages} {034022} (\bibinfo {year} {2018})},\ \Eprint
  {https://arxiv.org/abs/1709.05991} {arXiv:1709.05991 [hep-ph]} \BibitemShut
  {NoStop}%
\bibitem [{\citenamefont {Resch}\ \emph {et~al.}(2019)\citenamefont {Resch},
  \citenamefont {Rennecke},\ and\ \citenamefont {Schaefer}}]{Resch:2017vjs}%
  \BibitemOpen
  \bibfield  {author} {\bibinfo {author} {\bibfnamefont {S.}~\bibnamefont
  {Resch}}, \bibinfo {author} {\bibfnamefont {F.}~\bibnamefont {Rennecke}},\
  and\ \bibinfo {author} {\bibfnamefont {B.-J.}\ \bibnamefont {Schaefer}},\
  }\href {https://doi.org/10.1103/PhysRevD.99.076005} {\bibfield  {journal}
  {\bibinfo  {journal} {Phys. Rev. D}\ }\textbf {\bibinfo {volume} {99}},\
  \bibinfo {pages} {076005} (\bibinfo {year} {2019})},\ \Eprint
  {https://arxiv.org/abs/1712.07961} {arXiv:1712.07961 [hep-ph]} \BibitemShut
  {NoStop}%
\bibitem [{\citenamefont {Horn}\ and\ \citenamefont
  {Roberts}(2016)}]{Horn:2016rip}%
  \BibitemOpen
  \bibfield  {author} {\bibinfo {author} {\bibfnamefont {T.}~\bibnamefont
  {Horn}}\ and\ \bibinfo {author} {\bibfnamefont {C.~D.}\ \bibnamefont
  {Roberts}},\ }\href {https://doi.org/10.1088/0954-3899/43/7/073001}
  {\bibfield  {journal} {\bibinfo  {journal} {J. Phys. G}\ }\textbf {\bibinfo
  {volume} {43}},\ \bibinfo {pages} {073001} (\bibinfo {year} {2016})},\
  \Eprint {https://arxiv.org/abs/1602.04016} {arXiv:1602.04016 [nucl-th]}
  \BibitemShut {NoStop}%
\bibitem [{\citenamefont {Maris}\ \emph
  {et~al.}(1998{\natexlab{a}})\citenamefont {Maris}, \citenamefont {Roberts},\
  and\ \citenamefont {Schmidt}}]{Maris:1997eg}%
  \BibitemOpen
  \bibfield  {author} {\bibinfo {author} {\bibfnamefont {P.}~\bibnamefont
  {Maris}}, \bibinfo {author} {\bibfnamefont {C.~D.}\ \bibnamefont {Roberts}},\
  and\ \bibinfo {author} {\bibfnamefont {S.~M.}\ \bibnamefont {Schmidt}},\
  }\href {https://doi.org/10.1103/PhysRevC.57.R2821} {\bibfield  {journal}
  {\bibinfo  {journal} {Phys. Rev. C}\ }\textbf {\bibinfo {volume} {57}},\
  \bibinfo {pages} {R2821} (\bibinfo {year} {1998}{\natexlab{a}})},\ \Eprint
  {https://arxiv.org/abs/nucl-th/9801059} {arXiv:nucl-th/9801059 [nucl-th]}
  \BibitemShut {NoStop}%
\bibitem [{\citenamefont {Bender}\ \emph {et~al.}(1998)\citenamefont {Bender},
  \citenamefont {Poulis}, \citenamefont {Roberts}, \citenamefont {Schmidt},\
  and\ \citenamefont {Thomas}}]{Bender:1997jf}%
  \BibitemOpen
  \bibfield  {author} {\bibinfo {author} {\bibfnamefont {A.}~\bibnamefont
  {Bender}}, \bibinfo {author} {\bibfnamefont {G.~I.}\ \bibnamefont {Poulis}},
  \bibinfo {author} {\bibfnamefont {C.~D.}\ \bibnamefont {Roberts}}, \bibinfo
  {author} {\bibfnamefont {S.~M.}\ \bibnamefont {Schmidt}},\ and\ \bibinfo
  {author} {\bibfnamefont {A.~W.}\ \bibnamefont {Thomas}},\ }\href
  {https://doi.org/10.1016/S0370-2693(98)00546-2} {\bibfield  {journal}
  {\bibinfo  {journal} {Phys. Lett. B}\ }\textbf {\bibinfo {volume} {431}},\
  \bibinfo {pages} {263} (\bibinfo {year} {1998})},\ \Eprint
  {https://arxiv.org/abs/nucl-th/9710069} {arXiv:nucl-th/9710069 [nucl-th]}
  \BibitemShut {NoStop}%
\bibitem [{\citenamefont {Zong}\ \emph {et~al.}(2005)\citenamefont {Zong},
  \citenamefont {Chang}, \citenamefont {Hou}, \citenamefont {Sun},\ and\
  \citenamefont {Liu}}]{Zong:2005mm}%
  \BibitemOpen
  \bibfield  {author} {\bibinfo {author} {\bibfnamefont {H.-s.}\ \bibnamefont
  {Zong}}, \bibinfo {author} {\bibfnamefont {L.}~\bibnamefont {Chang}},
  \bibinfo {author} {\bibfnamefont {F.-y.}\ \bibnamefont {Hou}}, \bibinfo
  {author} {\bibfnamefont {W.-m.}\ \bibnamefont {Sun}},\ and\ \bibinfo {author}
  {\bibfnamefont {Y.-x.}\ \bibnamefont {Liu}},\ }\href
  {https://doi.org/10.1103/PhysRevC.71.015205} {\bibfield  {journal} {\bibinfo
  {journal} {Phys. Rev. C}\ }\textbf {\bibinfo {volume} {71}},\ \bibinfo
  {pages} {015205} (\bibinfo {year} {2005})}\BibitemShut {NoStop}%
\bibitem [{\citenamefont {Jiang}\ \emph
  {et~al.}(2008{\natexlab{a}})\citenamefont {Jiang}, \citenamefont {Shi},
  \citenamefont {Feng}, \citenamefont {Sun},\ and\ \citenamefont
  {Zong}}]{Jiang:2008zzd}%
  \BibitemOpen
  \bibfield  {author} {\bibinfo {author} {\bibfnamefont {Y.}~\bibnamefont
  {Jiang}}, \bibinfo {author} {\bibfnamefont {Y.-m.}\ \bibnamefont {Shi}},
  \bibinfo {author} {\bibfnamefont {H.-t.}\ \bibnamefont {Feng}}, \bibinfo
  {author} {\bibfnamefont {W.-m.}\ \bibnamefont {Sun}},\ and\ \bibinfo {author}
  {\bibfnamefont {H.-s.}\ \bibnamefont {Zong}},\ }\href
  {https://doi.org/10.1103/PhysRevC.78.025214} {\bibfield  {journal} {\bibinfo
  {journal} {Phys. Rev. C}\ }\textbf {\bibinfo {volume} {78}},\ \bibinfo
  {pages} {025214} (\bibinfo {year} {2008}{\natexlab{a}})}\BibitemShut
  {NoStop}%
\bibitem [{\citenamefont {Jiang}\ \emph
  {et~al.}(2008{\natexlab{b}})\citenamefont {Jiang}, \citenamefont {Shi},
  \citenamefont {Li}, \citenamefont {Sun},\ and\ \citenamefont
  {Zong}}]{Jiang:2008rb}%
  \BibitemOpen
  \bibfield  {author} {\bibinfo {author} {\bibfnamefont {Y.}~\bibnamefont
  {Jiang}}, \bibinfo {author} {\bibfnamefont {Y.-m.}\ \bibnamefont {Shi}},
  \bibinfo {author} {\bibfnamefont {H.}~\bibnamefont {Li}}, \bibinfo {author}
  {\bibfnamefont {W.-m.}\ \bibnamefont {Sun}},\ and\ \bibinfo {author}
  {\bibfnamefont {H.-s.}\ \bibnamefont {Zong}},\ }\href
  {https://doi.org/10.1103/PhysRevD.78.116005} {\bibfield  {journal} {\bibinfo
  {journal} {Phys. Rev. D}\ }\textbf {\bibinfo {volume} {78}},\ \bibinfo
  {pages} {116005} (\bibinfo {year} {2008}{\natexlab{b}})},\ \Eprint
  {https://arxiv.org/abs/0810.0750} {arXiv:0810.0750 [nucl-th]} \BibitemShut
  {NoStop}%
\bibitem [{\citenamefont {Eichmann}\ \emph
  {et~al.}(2016{\natexlab{a}})\citenamefont {Eichmann}, \citenamefont
  {Fischer},\ and\ \citenamefont {Welzbacher}}]{Eichmann:2015kfa}%
  \BibitemOpen
  \bibfield  {author} {\bibinfo {author} {\bibfnamefont {G.}~\bibnamefont
  {Eichmann}}, \bibinfo {author} {\bibfnamefont {C.~S.}\ \bibnamefont
  {Fischer}},\ and\ \bibinfo {author} {\bibfnamefont {C.~A.}\ \bibnamefont
  {Welzbacher}},\ }\href {https://doi.org/10.1103/PhysRevD.93.034013}
  {\bibfield  {journal} {\bibinfo  {journal} {Phys. Rev. D}\ }\textbf {\bibinfo
  {volume} {93}},\ \bibinfo {pages} {034013} (\bibinfo {year}
  {2016}{\natexlab{a}})},\ \Eprint {https://arxiv.org/abs/1509.02082}
  {arXiv:1509.02082 [hep-ph]} \BibitemShut {NoStop}%
\bibitem [{\citenamefont {Fischer}(2009)}]{Fischer:2009wc}%
  \BibitemOpen
  \bibfield  {author} {\bibinfo {author} {\bibfnamefont {C.~S.}\ \bibnamefont
  {Fischer}},\ }\href {https://doi.org/10.1103/PhysRevLett.103.052003}
  {\bibfield  {journal} {\bibinfo  {journal} {Phys. Rev. Lett.}\ }\textbf
  {\bibinfo {volume} {103}},\ \bibinfo {pages} {052003} (\bibinfo {year}
  {2009})},\ \Eprint {https://arxiv.org/abs/0904.2700} {arXiv:0904.2700
  [hep-ph]} \BibitemShut {NoStop}%
\bibitem [{\citenamefont {Fischer}\ \emph {et~al.}(2010)\citenamefont
  {Fischer}, \citenamefont {Maas},\ and\ \citenamefont
  {M\"uller}}]{Fischer:2010fx}%
  \BibitemOpen
  \bibfield  {author} {\bibinfo {author} {\bibfnamefont {C.~S.}\ \bibnamefont
  {Fischer}}, \bibinfo {author} {\bibfnamefont {A.}~\bibnamefont {Maas}},\ and\
  \bibinfo {author} {\bibfnamefont {J.~A.}\ \bibnamefont {M\"uller}},\ }\href
  {https://doi.org/10.1140/epjc/s10052-010-1343-1} {\bibfield  {journal}
  {\bibinfo  {journal} {Eur. Phys. J. C}\ }\textbf {\bibinfo {volume} {68}},\
  \bibinfo {pages} {165} (\bibinfo {year} {2010})},\ \Eprint
  {https://arxiv.org/abs/1003.1960} {arXiv:1003.1960 [hep-ph]} \BibitemShut
  {NoStop}%
\bibitem [{\citenamefont {Fischer}\ and\ \citenamefont
  {Mueller}(2011)}]{Fischer:2011pk}%
  \BibitemOpen
  \bibfield  {author} {\bibinfo {author} {\bibfnamefont {C.~S.}\ \bibnamefont
  {Fischer}}\ and\ \bibinfo {author} {\bibfnamefont {J.~A.}\ \bibnamefont
  {Mueller}},\ }\href {https://doi.org/10.1103/PhysRevD.84.054013} {\bibfield
  {journal} {\bibinfo  {journal} {Phys. Rev. D}\ }\textbf {\bibinfo {volume}
  {84}},\ \bibinfo {pages} {054013} (\bibinfo {year} {2011})},\ \Eprint
  {https://arxiv.org/abs/1106.2700} {arXiv:1106.2700 [hep-ph]} \BibitemShut
  {NoStop}%
\bibitem [{\citenamefont {Fischer}\ \emph {et~al.}(2011)\citenamefont
  {Fischer}, \citenamefont {Luecker},\ and\ \citenamefont
  {Mueller}}]{Fischer:2011mz}%
  \BibitemOpen
  \bibfield  {author} {\bibinfo {author} {\bibfnamefont {C.~S.}\ \bibnamefont
  {Fischer}}, \bibinfo {author} {\bibfnamefont {J.}~\bibnamefont {Luecker}},\
  and\ \bibinfo {author} {\bibfnamefont {J.~A.}\ \bibnamefont {Mueller}},\
  }\href {https://doi.org/10.1016/j.physletb.2011.07.039} {\bibfield  {journal}
  {\bibinfo  {journal} {Phys. Lett. B}\ }\textbf {\bibinfo {volume} {702}},\
  \bibinfo {pages} {438} (\bibinfo {year} {2011})},\ \Eprint
  {https://arxiv.org/abs/1104.1564} {arXiv:1104.1564 [hep-ph]} \BibitemShut
  {NoStop}%
\bibitem [{\citenamefont {Fischer}\ and\ \citenamefont
  {Luecker}(2013)}]{Fischer:2012vc}%
  \BibitemOpen
  \bibfield  {author} {\bibinfo {author} {\bibfnamefont {C.~S.}\ \bibnamefont
  {Fischer}}\ and\ \bibinfo {author} {\bibfnamefont {J.}~\bibnamefont
  {Luecker}},\ }\href {https://doi.org/10.1016/j.physletb.2012.11.054}
  {\bibfield  {journal} {\bibinfo  {journal} {Phys. Lett. B}\ }\textbf
  {\bibinfo {volume} {718}},\ \bibinfo {pages} {1036} (\bibinfo {year}
  {2013})},\ \Eprint {https://arxiv.org/abs/1206.5191} {arXiv:1206.5191
  [hep-ph]} \BibitemShut {NoStop}%
\bibitem [{\citenamefont {Fischer}\ \emph
  {et~al.}(2014{\natexlab{a}})\citenamefont {Fischer}, \citenamefont
  {Luecker},\ and\ \citenamefont {Welzbacher}}]{Fischer:2014ata}%
  \BibitemOpen
  \bibfield  {author} {\bibinfo {author} {\bibfnamefont {C.~S.}\ \bibnamefont
  {Fischer}}, \bibinfo {author} {\bibfnamefont {J.}~\bibnamefont {Luecker}},\
  and\ \bibinfo {author} {\bibfnamefont {C.~A.}\ \bibnamefont {Welzbacher}},\
  }\href {https://doi.org/10.1103/PhysRevD.90.034022} {\bibfield  {journal}
  {\bibinfo  {journal} {Phys. Rev. D}\ }\textbf {\bibinfo {volume} {90}},\
  \bibinfo {pages} {034022} (\bibinfo {year} {2014}{\natexlab{a}})},\ \Eprint
  {https://arxiv.org/abs/1405.4762} {arXiv:1405.4762 [hep-ph]} \BibitemShut
  {NoStop}%
\bibitem [{\citenamefont {M\"uller}\ \emph {et~al.}(2013)\citenamefont
  {M\"uller}, \citenamefont {Buballa},\ and\ \citenamefont
  {Wambach}}]{Muller:2013pya}%
  \BibitemOpen
  \bibfield  {author} {\bibinfo {author} {\bibfnamefont {D.}~\bibnamefont
  {M\"uller}}, \bibinfo {author} {\bibfnamefont {M.}~\bibnamefont {Buballa}},\
  and\ \bibinfo {author} {\bibfnamefont {J.}~\bibnamefont {Wambach}},\ }\href
  {https://doi.org/10.1140/epja/i2013-13096-5} {\bibfield  {journal} {\bibinfo
  {journal} {Eur. Phys. J. A}\ }\textbf {\bibinfo {volume} {49}},\ \bibinfo
  {pages} {96} (\bibinfo {year} {2013})},\ \Eprint
  {https://arxiv.org/abs/1303.2693} {arXiv:1303.2693 [hep-ph]} \BibitemShut
  {NoStop}%
\bibitem [{\citenamefont {M\"uller}\ \emph {et~al.}(2016)\citenamefont
  {M\"uller}, \citenamefont {Buballa},\ and\ \citenamefont
  {Wambach}}]{Muller:2016fdr}%
  \BibitemOpen
  \bibfield  {author} {\bibinfo {author} {\bibfnamefont {D.}~\bibnamefont
  {M\"uller}}, \bibinfo {author} {\bibfnamefont {M.}~\bibnamefont {Buballa}},\
  and\ \bibinfo {author} {\bibfnamefont {J.}~\bibnamefont {Wambach}},\
  }\href@noop {} {}\bibinfo {howpublished}
  {\href{http://arxiv.org/abs/1603.02865}{arXiv:1603.02865 [hep-ph]}} (\bibinfo
  {year} {2016})\BibitemShut {NoStop}%
\bibitem [{\citenamefont {Maas}\ \emph {et~al.}(2012)\citenamefont {Maas},
  \citenamefont {Pawlowski}, \citenamefont {von Smekal},\ and\ \citenamefont
  {Spielmann}}]{Maas:2011ez}%
  \BibitemOpen
  \bibfield  {author} {\bibinfo {author} {\bibfnamefont {A.}~\bibnamefont
  {Maas}}, \bibinfo {author} {\bibfnamefont {J.~M.}\ \bibnamefont {Pawlowski}},
  \bibinfo {author} {\bibfnamefont {L.}~\bibnamefont {von Smekal}},\ and\
  \bibinfo {author} {\bibfnamefont {D.}~\bibnamefont {Spielmann}},\ }\href
  {https://doi.org/10.1103/PhysRevD.85.034037} {\bibfield  {journal} {\bibinfo
  {journal} {Phys. Rev. D}\ }\textbf {\bibinfo {volume} {85}},\ \bibinfo
  {pages} {034037} (\bibinfo {year} {2012})},\ \Eprint
  {https://arxiv.org/abs/1110.6340} {arXiv:1110.6340 [hep-lat]} \BibitemShut
  {NoStop}%
\bibitem [{\citenamefont {Fischer}\ \emph
  {et~al.}(2014{\natexlab{b}})\citenamefont {Fischer}, \citenamefont {Fister},
  \citenamefont {L\"ucker},\ and\ \citenamefont {Pawlowski}}]{Fischer:2013eca}%
  \BibitemOpen
  \bibfield  {author} {\bibinfo {author} {\bibfnamefont {C.~S.}\ \bibnamefont
  {Fischer}}, \bibinfo {author} {\bibfnamefont {L.}~\bibnamefont {Fister}},
  \bibinfo {author} {\bibfnamefont {J.}~\bibnamefont {L\"ucker}},\ and\
  \bibinfo {author} {\bibfnamefont {J.~M.}\ \bibnamefont {Pawlowski}},\ }\href
  {https://doi.org/https://doi.org/10.1016/j.physletb.2014.03.057} {\bibfield
  {journal} {\bibinfo  {journal} {Phys. Lett. B}\ }\textbf {\bibinfo {volume}
  {732}},\ \bibinfo {pages} {273} (\bibinfo {year} {2014}{\natexlab{b}})},\
  \Eprint {https://arxiv.org/abs/1306.6022} {arXiv:1306.6022 [hep-lat]}
  \BibitemShut {NoStop}%
\bibitem [{\citenamefont {Aouane}\ \emph {et~al.}(2013)\citenamefont {Aouane},
  \citenamefont {Burger}, \citenamefont {Ilgenfritz}, \citenamefont
  {Müller-Preussker},\ and\ \citenamefont {Sternbeck}}]{Aouane:2012bk}%
  \BibitemOpen
  \bibfield  {author} {\bibinfo {author} {\bibfnamefont {R.}~\bibnamefont
  {Aouane}}, \bibinfo {author} {\bibfnamefont {F.}~\bibnamefont {Burger}},
  \bibinfo {author} {\bibfnamefont {E.~M.}\ \bibnamefont {Ilgenfritz}},
  \bibinfo {author} {\bibfnamefont {M.}~\bibnamefont {Müller-Preussker}},\
  and\ \bibinfo {author} {\bibfnamefont {A.}~\bibnamefont {Sternbeck}},\ }\href
  {https://doi.org/10.1103/PhysRevD.87.114502} {\bibfield  {journal} {\bibinfo
  {journal} {Phys. Rev. D}\ }\textbf {\bibinfo {volume} {87}},\ \bibinfo
  {pages} {114502} (\bibinfo {year} {2013})},\ \Eprint
  {https://arxiv.org/abs/1212.1102} {arXiv:1212.1102 [hep-lat]} \BibitemShut
  {NoStop}%
\bibitem [{\citenamefont {Taylor}(1971)}]{Taylor:1971ff}%
  \BibitemOpen
  \bibfield  {author} {\bibinfo {author} {\bibfnamefont {J.~C.}\ \bibnamefont
  {Taylor}},\ }\href {https://doi.org/10.1016/0550-3213(71)90297-5} {\bibfield
  {journal} {\bibinfo  {journal} {Nucl. Phys. B}\ }\textbf {\bibinfo {volume}
  {33}},\ \bibinfo {pages} {436} (\bibinfo {year} {1971})}\BibitemShut
  {NoStop}%
\bibitem [{\citenamefont {Isserstedt}\ \emph {et~al.}(2019)\citenamefont
  {Isserstedt}, \citenamefont {Buballa}, \citenamefont {Fischer},\ and\
  \citenamefont {Gunkel}}]{Isserstedt:2019pgx}%
  \BibitemOpen
  \bibfield  {author} {\bibinfo {author} {\bibfnamefont {P.}~\bibnamefont
  {Isserstedt}}, \bibinfo {author} {\bibfnamefont {M.}~\bibnamefont {Buballa}},
  \bibinfo {author} {\bibfnamefont {C.~S.}\ \bibnamefont {Fischer}},\ and\
  \bibinfo {author} {\bibfnamefont {P.~J.}\ \bibnamefont {Gunkel}},\
  }\href@noop {} {}\bibinfo {howpublished}
  {\href{http://arxiv.org/abs/1906.11644}{arXiv:1906.11644 [hep-ph]}} (\bibinfo
  {year} {2019})\BibitemShut {NoStop}%
\bibitem [{\citenamefont {Braun}\ \emph {et~al.}(2016)\citenamefont {Braun},
  \citenamefont {Fister}, \citenamefont {Pawlowski},\ and\ \citenamefont
  {Rennecke}}]{Braun:2014ata}%
  \BibitemOpen
  \bibfield  {author} {\bibinfo {author} {\bibfnamefont {J.}~\bibnamefont
  {Braun}}, \bibinfo {author} {\bibfnamefont {L.}~\bibnamefont {Fister}},
  \bibinfo {author} {\bibfnamefont {J.~M.}\ \bibnamefont {Pawlowski}},\ and\
  \bibinfo {author} {\bibfnamefont {F.}~\bibnamefont {Rennecke}},\ }\href
  {https://doi.org/10.1103/PhysRevD.94.034016} {\bibfield  {journal} {\bibinfo
  {journal} {Phys. Rev. D}\ }\textbf {\bibinfo {volume} {94}},\ \bibinfo
  {pages} {034016} (\bibinfo {year} {2016})},\ \Eprint
  {https://arxiv.org/abs/1412.1045} {arXiv:1412.1045 [hep-ph]} \BibitemShut
  {NoStop}%
\bibitem [{\citenamefont {Williams}\ \emph {et~al.}(2016)\citenamefont
  {Williams}, \citenamefont {Fischer},\ and\ \citenamefont
  {Heupel}}]{Williams:2015cvx}%
  \BibitemOpen
  \bibfield  {author} {\bibinfo {author} {\bibfnamefont {R.}~\bibnamefont
  {Williams}}, \bibinfo {author} {\bibfnamefont {C.~S.}\ \bibnamefont
  {Fischer}},\ and\ \bibinfo {author} {\bibfnamefont {W.}~\bibnamefont
  {Heupel}},\ }\href {https://doi.org/10.1103/PhysRevD.93.034026} {\bibfield
  {journal} {\bibinfo  {journal} {Phys. Rev. D}\ }\textbf {\bibinfo {volume}
  {93}},\ \bibinfo {pages} {034026} (\bibinfo {year} {2016})},\ \Eprint
  {https://arxiv.org/abs/1512.00455} {arXiv:1512.00455 [hep-ph]} \BibitemShut
  {NoStop}%
\bibitem [{\citenamefont {Ball}\ and\ \citenamefont
  {Chiu}(1980)}]{Ball:1980ay}%
  \BibitemOpen
  \bibfield  {author} {\bibinfo {author} {\bibfnamefont {J.~S.}\ \bibnamefont
  {Ball}}\ and\ \bibinfo {author} {\bibfnamefont {T.-W.}\ \bibnamefont
  {Chiu}},\ }\href {https://doi.org/10.1103/PhysRevD.22.2542} {\bibfield
  {journal} {\bibinfo  {journal} {Phys. Rev. D}\ }\textbf {\bibinfo {volume}
  {22}},\ \bibinfo {pages} {2542} (\bibinfo {year} {1980})}\BibitemShut
  {NoStop}%
\bibitem [{\citenamefont {Heupel}\ \emph {et~al.}(2014)\citenamefont {Heupel},
  \citenamefont {Goecke},\ and\ \citenamefont {Fischer}}]{Heupel:2014ina}%
  \BibitemOpen
  \bibfield  {author} {\bibinfo {author} {\bibfnamefont {W.}~\bibnamefont
  {Heupel}}, \bibinfo {author} {\bibfnamefont {T.}~\bibnamefont {Goecke}},\
  and\ \bibinfo {author} {\bibfnamefont {C.~S.}\ \bibnamefont {Fischer}},\
  }\href {https://doi.org/10.1140/epja/i2014-14085-x} {\bibfield  {journal}
  {\bibinfo  {journal} {Eur. Phys. J. A}\ }\textbf {\bibinfo {volume} {50}},\
  \bibinfo {pages} {85} (\bibinfo {year} {2014})},\ \Eprint
  {https://arxiv.org/abs/1402.5042} {arXiv:1402.5042 [hep-ph]} \BibitemShut
  {NoStop}%
\bibitem [{\citenamefont {Bellwied}\ \emph {et~al.}(2015)\citenamefont
  {Bellwied} \emph {et~al.}}]{Bellwied:2015rza}%
  \BibitemOpen
  \bibfield  {author} {\bibinfo {author} {\bibfnamefont {R.}~\bibnamefont
  {Bellwied}} \emph {et~al.},\ }\href
  {https://doi.org/10.1016/j.physletb.2015.11.011} {\bibfield  {journal}
  {\bibinfo  {journal} {Phys. Lett. B}\ }\textbf {\bibinfo {volume} {751}},\
  \bibinfo {pages} {559} (\bibinfo {year} {2015})},\ \Eprint
  {https://arxiv.org/abs/1507.07510} {arXiv:1507.07510 [hep-lat]} \BibitemShut
  {NoStop}%
\bibitem [{\citenamefont {Bazavov}\ \emph {et~al.}(2019)\citenamefont {Bazavov}
  \emph {et~al.}}]{Bazavov:2018mes}%
  \BibitemOpen
  \bibfield  {author} {\bibinfo {author} {\bibfnamefont {A.}~\bibnamefont
  {Bazavov}} \emph {et~al.},\ }\href
  {https://doi.org/10.1016/j.physletb.2019.05.013} {\bibfield  {journal}
  {\bibinfo  {journal} {Phys. Lett. B}\ }\textbf {\bibinfo {volume} {795}},\
  \bibinfo {pages} {15} (\bibinfo {year} {2019})},\ \Eprint
  {https://arxiv.org/abs/1812.08235} {arXiv:1812.08235 [hep-lat]} \BibitemShut
  {NoStop}%
\bibitem [{\citenamefont {Sanchis-Alepuz}\ and\ \citenamefont
  {Williams}(2018)}]{Sanchis-Alepuz:2017jjd}%
  \BibitemOpen
  \bibfield  {author} {\bibinfo {author} {\bibfnamefont {H.}~\bibnamefont
  {Sanchis-Alepuz}}\ and\ \bibinfo {author} {\bibfnamefont {R.}~\bibnamefont
  {Williams}},\ }\href
  {https://doi.org/https://doi.org/10.1016/j.cpc.2018.05.020} {\bibfield
  {journal} {\bibinfo  {journal} {Comput. Phys. Commun.}\ }\textbf {\bibinfo
  {volume} {232}},\ \bibinfo {pages} {1} (\bibinfo {year} {2018})},\ \Eprint
  {https://arxiv.org/abs/1710.04903} {arXiv:1710.04903 [hep-ph]} \BibitemShut
  {NoStop}%
\bibitem [{\citenamefont {Eichmann}\ \emph
  {et~al.}(2016{\natexlab{b}})\citenamefont {Eichmann}, \citenamefont
  {Sanchis-Alepuz}, \citenamefont {Williams}, \citenamefont {Alkofer},\ and\
  \citenamefont {Fischer}}]{Eichmann:2016yit}%
  \BibitemOpen
  \bibfield  {author} {\bibinfo {author} {\bibfnamefont {G.}~\bibnamefont
  {Eichmann}}, \bibinfo {author} {\bibfnamefont {H.}~\bibnamefont
  {Sanchis-Alepuz}}, \bibinfo {author} {\bibfnamefont {R.}~\bibnamefont
  {Williams}}, \bibinfo {author} {\bibfnamefont {R.}~\bibnamefont {Alkofer}},\
  and\ \bibinfo {author} {\bibfnamefont {C.~S.}\ \bibnamefont {Fischer}},\
  }\href {https://doi.org/10.1016/j.ppnp.2016.07.001} {\bibfield  {journal}
  {\bibinfo  {journal} {Prog. Part. Nucl. Phys.}\ }\textbf {\bibinfo {volume}
  {91}},\ \bibinfo {pages} {1} (\bibinfo {year} {2016}{\natexlab{b}})},\
  \Eprint {https://arxiv.org/abs/1606.09602} {arXiv:1606.09602 [hep-ph]}
  \BibitemShut {NoStop}%
\bibitem [{\citenamefont {Nakanishi}(1965)}]{Nakanishi:1965zza}%
  \BibitemOpen
  \bibfield  {author} {\bibinfo {author} {\bibfnamefont {N.}~\bibnamefont
  {Nakanishi}},\ }\href {https://doi.org/10.1103/PhysRev.138.B1182} {\bibfield
  {journal} {\bibinfo  {journal} {Phys. Rev.}\ }\textbf {\bibinfo {volume}
  {138}},\ \bibinfo {pages} {B1182} (\bibinfo {year} {1965})}\BibitemShut
  {NoStop}%
\bibitem [{\citenamefont {Son}\ and\ \citenamefont
  {Stephanov}(2002{\natexlab{a}})}]{Son:2001ff}%
  \BibitemOpen
  \bibfield  {author} {\bibinfo {author} {\bibfnamefont {D.~T.}\ \bibnamefont
  {Son}}\ and\ \bibinfo {author} {\bibfnamefont {M.~A.}\ \bibnamefont
  {Stephanov}},\ }\href {https://doi.org/10.1103/PhysRevLett.88.202302}
  {\bibfield  {journal} {\bibinfo  {journal} {Phys. Rev. Lett.}\ }\textbf
  {\bibinfo {volume} {88}},\ \bibinfo {pages} {202302} (\bibinfo {year}
  {2002}{\natexlab{a}})},\ \Eprint {https://arxiv.org/abs/0111100}
  {arXiv:0111100 [hep-ph]} \BibitemShut {NoStop}%
\bibitem [{\citenamefont {Son}\ and\ \citenamefont
  {Stephanov}(2002{\natexlab{b}})}]{Son:2002ci}%
  \BibitemOpen
  \bibfield  {author} {\bibinfo {author} {\bibfnamefont {D.~T.}\ \bibnamefont
  {Son}}\ and\ \bibinfo {author} {\bibfnamefont {M.~A.}\ \bibnamefont
  {Stephanov}},\ }\href {https://doi.org/10.1103/PhysRevD.66.076011} {\bibfield
   {journal} {\bibinfo  {journal} {Phys. Rev. D}\ }\textbf {\bibinfo {volume}
  {66}},\ \bibinfo {pages} {076011} (\bibinfo {year} {2002}{\natexlab{b}})},\
  \Eprint {https://arxiv.org/abs/0204226} {arXiv:0204226 [hep-ph]} \BibitemShut
  {NoStop}%
\bibitem [{\citenamefont {Williams}\ \emph {et~al.}(2007)\citenamefont
  {Williams}, \citenamefont {Fischer},\ and\ \citenamefont
  {Pennington}}]{Williams:2006vva}%
  \BibitemOpen
  \bibfield  {author} {\bibinfo {author} {\bibfnamefont {R.}~\bibnamefont
  {Williams}}, \bibinfo {author} {\bibfnamefont {C.~S.}\ \bibnamefont
  {Fischer}},\ and\ \bibinfo {author} {\bibfnamefont {M.~R.}\ \bibnamefont
  {Pennington}},\ }\href {https://doi.org/10.1016/j.physletb.2006.12.055}
  {\bibfield  {journal} {\bibinfo  {journal} {Phys. Lett. B}\ }\textbf
  {\bibinfo {volume} {645}},\ \bibinfo {pages} {167} (\bibinfo {year}
  {2007})},\ \Eprint {https://arxiv.org/abs/hep-ph/0612061}
  {arXiv:hep-ph/0612061 [hep-ph]} \BibitemShut {NoStop}%
\bibitem [{\citenamefont {Chang}\ \emph {et~al.}(2007)\citenamefont {Chang},
  \citenamefont {Liu}, \citenamefont {Bhagwat}, \citenamefont {Roberts},\ and\
  \citenamefont {Wright}}]{Chang:2006bm}%
  \BibitemOpen
  \bibfield  {author} {\bibinfo {author} {\bibfnamefont {L.}~\bibnamefont
  {Chang}}, \bibinfo {author} {\bibfnamefont {Y.-X.}\ \bibnamefont {Liu}},
  \bibinfo {author} {\bibfnamefont {M.~S.}\ \bibnamefont {Bhagwat}}, \bibinfo
  {author} {\bibfnamefont {C.~D.}\ \bibnamefont {Roberts}},\ and\ \bibinfo
  {author} {\bibfnamefont {S.~V.}\ \bibnamefont {Wright}},\ }\href
  {https://doi.org/10.1103/PhysRevC.75.015201} {\bibfield  {journal} {\bibinfo
  {journal} {Phys. Rev. C}\ }\textbf {\bibinfo {volume} {75}},\ \bibinfo
  {pages} {015201} (\bibinfo {year} {2007})},\ \Eprint
  {https://arxiv.org/abs/nucl-th/0605058} {arXiv:nucl-th/0605058 [nucl-th]}
  \BibitemShut {NoStop}%
\bibitem [{\citenamefont {Fischer}\ \emph {et~al.}(2009)\citenamefont
  {Fischer}, \citenamefont {Nickel},\ and\ \citenamefont
  {Williams}}]{Fischer:2008sp}%
  \BibitemOpen
  \bibfield  {author} {\bibinfo {author} {\bibfnamefont {C.~S.}\ \bibnamefont
  {Fischer}}, \bibinfo {author} {\bibfnamefont {D.}~\bibnamefont {Nickel}},\
  and\ \bibinfo {author} {\bibfnamefont {R.}~\bibnamefont {Williams}},\ }\href
  {https://doi.org/10.1140/epjc/s10052-008-0821-1} {\bibfield  {journal}
  {\bibinfo  {journal} {Eur. Phys. J. C}\ }\textbf {\bibinfo {volume} {60}},\
  \bibinfo {pages} {47} (\bibinfo {year} {2009})},\ \Eprint
  {https://arxiv.org/abs/0807.3486} {arXiv:0807.3486 [hep-ph]} \BibitemShut
  {NoStop}%
\bibitem [{\citenamefont {Maris}\ \emph
  {et~al.}(1998{\natexlab{b}})\citenamefont {Maris}, \citenamefont {Roberts},\
  and\ \citenamefont {Tandy}}]{Maris:1997hd}%
  \BibitemOpen
  \bibfield  {author} {\bibinfo {author} {\bibfnamefont {P.}~\bibnamefont
  {Maris}}, \bibinfo {author} {\bibfnamefont {C.~D.}\ \bibnamefont {Roberts}},\
  and\ \bibinfo {author} {\bibfnamefont {P.~C.}\ \bibnamefont {Tandy}},\ }\href
  {https://doi.org/10.1016/S0370-2693(97)01535-9} {\bibfield  {journal}
  {\bibinfo  {journal} {Phys. Lett. B}\ }\textbf {\bibinfo {volume} {420}},\
  \bibinfo {pages} {267} (\bibinfo {year} {1998}{\natexlab{b}})},\ \Eprint
  {https://arxiv.org/abs/nucl-th/9707003} {arXiv:nucl-th/9707003 [nucl-th]}
  \BibitemShut {NoStop}%
\bibitem [{\citenamefont {Detmold}(2002)}]{Detmold:2002}%
  \BibitemOpen
  \bibfield  {author} {\bibinfo {author} {\bibfnamefont {W.}~\bibnamefont
  {Detmold}},\ }\emph {\bibinfo {title} {{Nonperturbative approaches to quantum
  chromodynamics}}},\ \href
  {https://digital.library.adelaide.edu.au/dspace/handle/2440/21733} {Ph.D.
  thesis},\ \bibinfo  {school} {University of Adelaide}, \bibinfo {address}
  {Australia} (\bibinfo {year} {2002})\BibitemShut {NoStop}%
\end{thebibliography}%

\end{document}